\documentclass[a4paper,fleqn,usenatbib]{mnras}

\usepackage[T1]{fontenc}
\usepackage{ae,aecompl}

\usepackage{graphicx}	% Including figure files
\usepackage{amsmath}	% Advanced maths commands
\usepackage{amssymb}	% Extra maths symbols
\usepackage{array}
\usepackage{multirow}
\usepackage{esint}
\usepackage{natbib}
\usepackage{hyperref}
\usepackage{cleveref}
\usepackage{verbatim}        % for debugging and notes
\usepackage{color}

\newcommand{\nway}{\textsc{Nway\,}}
\newcommand{\code}[1]{{\texttt{#1}}}
\newcommand{\xmm}{\textit{XMM-Newton}}

\newcommand{\am}{\text{$^\prime$}}
\newcommand{\as}{\text{$^{\prime\prime}$}}
\newcommand{\cgs}{\text{erg\,cm$^{-2}$\,s$^{-1}$}}

\title[AllWISE counterparts to ROSAT and XMMSL2 sources]{Finding counterparts for all-sky X-ray surveys with \textsc{Nway}: 
a Bayesian algorithm for cross-matching multiple catalogues }

\author[M. Salvato et al.]{
Mara Salvato,$^{1,2}$\thanks{E-mail: mara@mpe.mpg.de (MS)}
J. Buchner,$^{3,4}$
T. Budav\'ari,$^{5}$
T. Dwelly,$^{1}$
A. Merloni,$^{1,2}$
\newauthor
M. Brusa,$^{6,7}$
A. Rau,$^{1}$
S. Fotopoulou$^{8}$
and K. Nandra$^{1,9}$
\\
$^{1}$MPE, Giessenbachstrasse 1, Garching 85748, Germany\\
$^{2}$Cluster of Excellence, Boltzmann Strasse 2, 85748, Germany\\
$^{3}$Pontificia Universidad Católica de Chile, Instituto de Astrofísica,
Casilla 306, Santiago 22, Chile\\
$^{4}$Millenium Institute of Astrophysics, Vicu\~{n}a. MacKenna 4860, 7820436
Macul, Santiago, Chile\\
$^{5}$The Johns Hopkins University, 3400 N. Charles Street, Baltimore, MD 21218, USA\\
$^{6}$Dipartimento di Fisica e Astronomia, Universit\'a di Bologna, Via Gobetti 93/2, 40129 Bologna, 
Italy\\
$^{7}$INAF- Osservatorio Astronomico di Bologna, Via Gobetti 93/3, 40129 Bologna, Italy\\
$^{8}$Department of Astronomy at the University of Geneva, Chemin d'Ecogia 16, 1290 Versoix, Switzerland\\
$^{9}$Imperial college, Kensington, London SW7 2AZ, UK
}

\date{Accepted XXX. Received YYY; in original form ZZZ}

\pubyear{2017}

\begin{document}
\label{firstpage}
\pagerange{\pageref{firstpage}--\pageref{lastpage}}
\maketitle

\begin{abstract}
  
We release the AllWISE counterparts and {\it Gaia} matches to 106,573 and 17,665 X-ray sources detected in 
the ROSAT 2RXS and XMMSL2 surveys with |b|$>$15$^\circ$. These are the brightest X-ray sources in the sky, but 
their position uncertainties and the sparse multi-wavelength coverage until now rendered the 
identification of their counterparts a demanding task with uncertain results. New all-sky multi-wavelength
surveys of sufficient depth, like AllWISE and {\it Gaia}, and a new Bayesian statistics based algorithm,
 \nway, allow us, for the first time, to provide reliable counterpart associations. \nway extends previous 
 distance and sky density based association methods and, using one or  more priors (e.g., colors, magnitudes), weights 
 the probability that sources from two or more catalogues  are simultaneously associated on the basis of 
 their  observable characteristics. Here, counterparts have been determined using a WISE color-magnitude 
 prior. A reference sample of 4524 XMM/{\it Chandra} and Swift X-ray sources demonstrates a 
 reliability of $\sim$ 94.7\% (2RXS) and 97.4\% (XMMSL2). Combining our results with {\it Chandra}-COSMOS
  data, we propose a new separation between stars and AGN in the X-ray/WISE flux-magnitude plane, valid over six orders of magnitude. 
We also release the \nway code and its user manual.
\nway was extensively tested with XMM-COSMOS data. Using two different sets of priors, we find an 
agreement of 96\,\% and 99\,\% with published Likelihood Ratio methods. Our results were achieved faster 
and without any follow-up visual inspection. With the advent of deep and wide area surveys in  X-rays 
(e.g. SRG/eROSITA, Athena/WFI) and radio  (ASKAP/EMU, LOFAR, APERTIF, etc.) \nway will provide a powerful and 
reliable counterpart identification tool.

\end{abstract}


%%%%%%%%%%%%%%%%%%%%%%%%%%%%%%%%%%%%%%%%%%%%%%%%%%

%%%%%%%%%%%%%%%%% BODY OF PAPER %%%%%%%%%%%%%%%%%%

\section{Introduction}

Active Galactic Nuclei (AGN) play an important role in the evolution of galaxies in the Universe. It is 
now established that  most massive galaxies host a supermassive black hole in their centre, and that the 
black hole accretion activity and history might have a profound influence on their growth. A 
comprehensive picture of this link can only be obtained from a complete census of AGN, covering the full 
luminosity function at any redshift. This is possible solely by merging AGN samples selected at 
different wavelengths and through complementary criteria \citep{Padovani:2017qy}, and by combining shallow wide-area with deep 
pencil beam surveys. The broad wavelength coverage is required to identify AGN at all redshifts at the 
wavelengths where they dominate the Spectral Energy Distribution (SED) of their host galaxy (e.g., Gamma-ray: 
\citealt{Armstrong:2015cr}; X-ray: \citealt{Georgakakis:2011dq}; optical: \citealt{Bovy:2011qy, Palanque-Delabrouille:2016nx};
Mid-Infrared: \citealt{Assef:2013sp};  Radio: \citealt{De-Breuck:2002bh}). Pencil beam 
surveys \citep[e.g.,][]{Luo:2017fk} allow the study of the high-redshift population and the faint end of 
the luminosity distribution, while shallower wide-area surveys \citep[e.g.,][]{Georgakakis:2017ys, 
LaMassa:2016cq} trace the brightest sources and at the same time provide access to rare objects.

The selection of AGN at X-ray energies provides an excellent compromise between completeness and purity 
of the sample. X-rays are sensitive to all but the most obscured AGN even when hosted in luminous 
galaxies and have very low contamination from other source populations. Limited by the available 
datasets, and by the small field of view of the most sensitive imaging telescopes, X-ray selected AGN 
samples were until now predominantly obtained from deep pencil beam surveys (e.g., COSMOS: 
\citealt{Hasinger:2007dn, Brusa:2010lr, Civano:2012ys, Marchesi:2016jw};  CDFS: \citealt{Luo:2010qp, Hsu:2014fj, 
Luo:2017fk}; AEGIS-X: \citealt{Nandra:2015uq}; Lockman Hole: \citealt{Fotopoulou:2012kx}) or limited 
to the brightest and most extreme sources \citep[e.g., BAT:][]{Baumgartner:2013vn}. Only very recently 
Stripe82X \citep[][]{LaMassa:2016cq, Tasnim-Ananna:2017gf} and XMM-XXL \citep[e.g.,][]{Pierre:2017zr, 
Fotopoulou:2016ly, Georgakakis:2017ys} opened access to two shallow, wide areas of $\approx$30\,deg$^{2}$ and $\approx$50\,deg$^{2}$. Still, the total population of X-ray selected and spectroscopically confirmed AGN counts only
 $\approx20,000$ objects and continues to be dwarfed by the  $\approx300,000$ optically selected quasars 
\citep[e.g., DR12Q:][]{Paris:2017qf}. The new revisions of the ROSAT All-sky Survey \citep[2RXS;][]
{Boller:2016qy}  and the second release of the XMM-Newton Slew Survey (XMMSL2\footnote{\url{https://
www.cosmos.esa.int/web/xmm-newton/xmmsl2-ug}}) with a total of $\approx130,000$ X-ray sources may finally 
provide AGN counts comparable to those found in the SDSS. 

So far, the most challenging aspect of the exploitation of these samples was  the identification of the 
multi-wavelength counterparts needed for the source characterization and redshift estimates. This was 
related to at least two shortcomings. First, the positional uncertainties of all but the brightest sources in 
these X-ray catalogues  are in general too large to assign a single, unambiguous optical counterpart 
based solely on a simple coordinate match. Second, the multi-wavelength catalogues  used for 
identifying the counterparts lacked depth and homogeneous, contiguous coverage. At least the latter 
problem can now be addressed  with the publicly available AllWISE survey \citep[i.e. the combination of WISE and NEOWISE:][] 
{Wright:2010yq, Mainzer:2011lh, Mainzer:2014mw}. This survey maps the entire 
sky at mid-infrared wavelengths from 3.4 to 22\,$\mu$m to a depth at which the majority of the point-source
 populations of 2RXS and XMMSL2 (AGN, stars, star-forming galaxies) is expected to be detected\footnote{Note, that the detection of an AGN in the mid-infrared requires the availability of 
reprocessing dust, i.e. dust free AGN will be missed. Many Compton-Thick AGN will be missed as well.} (see 
\S~\ref{sec:prior1}).

Even with a suitable catalogue at hand, the large X-ray positional uncertainties still require us to 
recognize the  correct counterpart among the many that are possible, avoiding the bias toward the optically brightest sources \citep[e.g.,][]{Naylor:2013fr}. The most frequently used technique is 
based on  the Likelihood Ratio (LR) method \citep{Sutherland:1992fv}. Using a primary catalogue
(here X-rays) and a secondary catalogue (here mid-infrared) the ratio of the likelihoods of each IR source being 
the true counterpart to a X-ray  or  background source is calculated taking into account the 
coordinates (i.e. their separations), the associated uncertainties, the density of the sources in the two 
catalogues, and the source magnitudes and  distribution. For X-ray sources with large positional 
uncertainties this limited set of information is often insufficient to reliably identify the 
counterpart.

For this reason we developed a new code, \nway, that goes beyond the LR approach by 
simultaneously considering in addition to astrometric information (i.e. position, associated uncertainties 
and sky density of sources as a function of magnitude), various known source properties (e.g. magnitudes, colours) using Bayesian statistics for each step.

The paper focusses on two main topics: firstly, we increase the sample of bright X-ray selected AGN 
by identifying and releasing the coordinates  of the AllWISE counterparts to the point-like X-ray 
sources in 2RXS and XMMSL2 all-sky surveys. This will facilitate spectroscopic follow-up and further 
source characterization \citep[e.g.,][]{Dwelly:2017qy}. Secondly, we present the \nway code and 
release it to the public, together with a detailed user manual. In order to keep the two aspects 
separated, in the main body of the paper we only provide a short description 
of \nway (\S~\ref{sec:application}). Instead we focus on the X-ray catalogues  (\S~\ref{sec:datasets}),
the construction of the prior based on AllWISE photometry (\S~\ref{sec:prior}), the assessment of the reliability of our associations by comparison with the literature (\S~\ref{sec:literature}), and the AllWISE properties of the counterparts (\S~\ref{sec:characterization}), in comparison with the results from X-ray pencil-beam surveys. The release of the catalogues  is 
presented in \S~\ref{sec:catalogues}.
The detailed description of the \nway algorithm and the verification results are made available in the 
Appendixes~\ref{A:intro} and \ref{A:nway}. Test performances of \nway are presented in Appendix~\ref{A:TestCosmos},
where we also show the strength of the method and the improvement of simultaneously using two priors instead of one.
 
Along the paper we assume Vega magnitudes unless differently stated. In order to
allow direct comparison with existing works from the literature of X-ray surveys, we adopt a flat 
$\Lambda$CDM cosmology with $h=H_0/[100 \mathrm{km s}^{-1} \mathrm{Mpc}^{-1}]=0.7$; $\Omega_M$=0.3; $\Omega_M$=0.7.

%---------------------------------------------------------------------
\section{The Datasets}
\label{sec:datasets}

In the following we describe the properties of the 2RXS, XMMSL2, and AllWISE catalogues  and their 
preparation for this work.

\subsection{ROSAT All-Sky Survey}
\label{subsec:2rxs}

The first all-sky imaging X-ray survey in the 0.1-2.4 keV band was performed by ROSAT \citep{Truemper:1982fu}
 between 1990 and 1991. Besides a catalogue of extended sources, two catalogues  of point-like 
sources were published: the Bright Source Catalogue (BSC) containing the 18,816 brightest sources 
\citep{Voges:1999fk} and the Faint Source Catalogue (FSC) encompassing the 105,924 fainter  
objects down to a detection likelihood limit of 6.5 \citep{Voges:2000qy}. In view of the launch of 
SRG/eROSITA \citep{Merloni:2012uq} and taking advantage of the advancement in technology, data reduction, 
analysis, and detection algorithms of the last 25 years, the original data have recently been reprocessed 
by \citet{Boller:2016qy}. The newly generated catalogue (2RXS) for point-like X-ray sources has 
been released to the community\footnote{\url{http://www.mpe.mpg.de/ROSATcompared with/2RXS}} and includes 
$\approx$13,5000 sources. 
 
\begin{figure*}
\centering
\includegraphics[width=8.0cm]{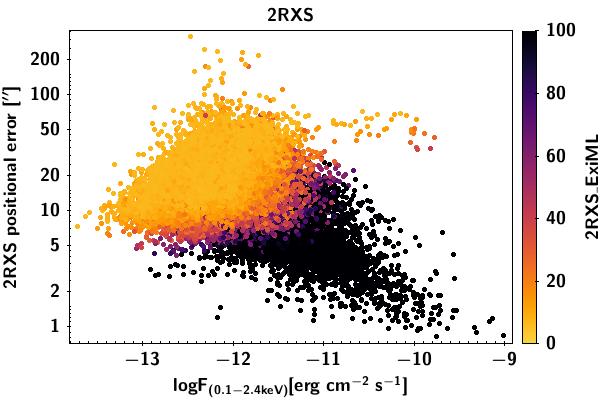}
\hspace*{1.0cm}
\includegraphics[width=8.0cm]{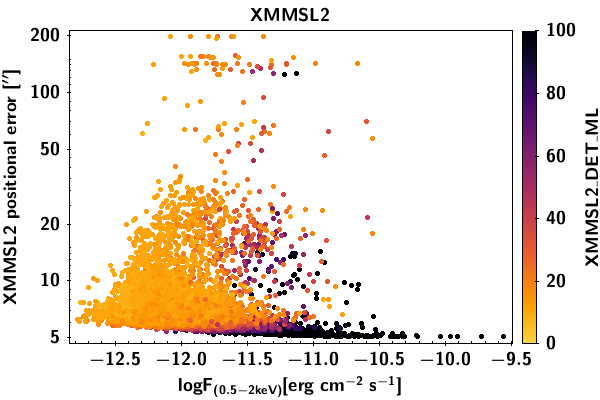}
     \caption{Positional uncertainties for the 2RXS (left)  and XMMSL2 (right) samples  as a function of 
X-ray flux , with the sources colour coded on the basis of their respective detection likelihood. The flux of the XMMSL2 sources in the 0.2-12 keV band has been converted into the 0.5-2 keV 
band assuming Galactic N$_{H}$ =3e20 cm$^{-2}$ and a power-law of 1.7. For the 802  XMMSL2 sources 
 without cataloged 0.2-12 keV flux we converted either the flux from the 0.2-2 keV band (775 sources) or the flux from
the 2-12 keV band (27 sources). 
\label{fig:2RXSproperties}}     
\end{figure*}

When comparing with the 1RXS catalog, which combines BSC and FSC, the number of reliable sources in the 2RXS has 
increased (both bright and faint) while the number of spurious detections has decreased \citep[see][for 
more details]{Boller:2016qy}.
We select all 2RXS detections  which lie within the `extragalactic' part of the sky, i.e. with $|b|>15^\circ$
and at least 6 and 3\,deg away from the optical centres of the Large and Small Magellanic Clouds, 
respectively. After this geometric filter, we are left with 106695 2RXS X-ray detections  with an 
estimated coverage of 30,575.9 deg$^2$.
Observed in projection outside the crowded Galactic Plane, these sources are predominantly 
extragalactic. The catalogue is further cleaned by removing 122 sources without estimated positional 
uncertainty and without listed counts.
The well known correlation between X-ray flux\footnote{We computed Galactic foreground absorption 
corrected fluxes following the procedure presented in Appendix A of \citet{Dwelly:2017qy}} intensity, 
positional uncertainty and detection likelihood is shown for the final  106,573 sources in the primary 
catalogue in the left panel of Fig.~\ref{fig:2RXSproperties}, with the flux distribution (converted into 
the 0.5-2 keV band) shown in Fig.~\ref{fig:flux_distribution}. 95\,\%  of the sources have a  1$\sigma$ 
positional error smaller than 29$\as$ compared with the 34$\as$ found in  the extragalactic area of 1RXS.

%------------------------------------------------------------------

%------------------------------------------------------------------

\subsection{\xmm\ Slew 2 survey}
\label{subsec:xmmsl2}

The \xmm\ European Photon Imaging Camera pn (EPIC-pn) accumulates data during slews between 
pointed observations. The most recent catalogue derived from this dataset covers 84\,\% of the sky 
(release~2.0, 14th March 2017). We extracted all detections from the `Clean' version of the catalogue 
(which we will henceforth refer to as the XMMSL2 catalogue), which lie in the same area  as defined for 
2RXS. After this geometric filter, we are left with 22,306 X-ray detections with at least 0.1\,s of 
effective XMMSL2 exposure with an estimated coverage of  $\approx$25,500\,deg$^2$.

The final catalogue was filtered to remove candidate duplicate detections of identical X-ray sources 
using the original column UNIQUE\_SRCNAME, retaining a total of  17,672 sources with  2,704 sources detected 
only at 0.2-12 keV, 553 detected only at 0.2-2 keV , and 168 sources detected only at 2-12 keV.

\begin{figure}
\centering
\includegraphics[width=8.0cm]{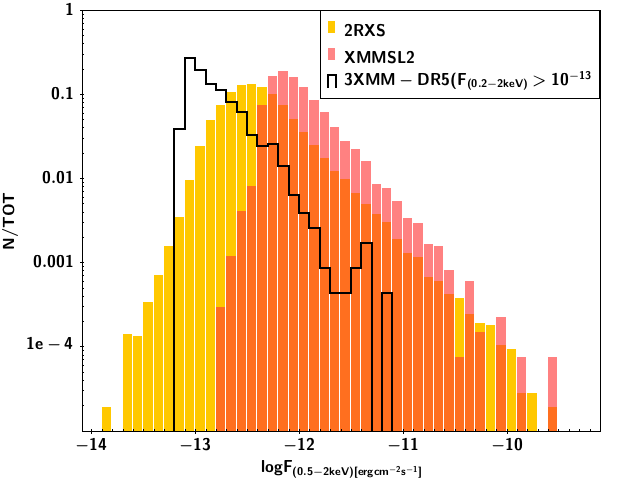}
\caption{Flux distribution for the 2RXS (yellow), XMMSL2 (brown) and 3XMM-DR5 catalogues. The flux from 
the original bands  has been transformed to the flux at (0.5-2 keV), assuming a Galactic N$_{H}$ 
=3(2.29)e20 cm$^{-2}$ and a power-law of 1.7(2.4) for XMM SL2(2RXS) data, respectively. 
     \label{fig:flux_distribution}}     
\end{figure}

$52.8$\,\% ( 9,333) of the XMMSL2 sources have at least one 2RXS source within a radius of 60$\as$, with 
236/21/3/1/1 XMMSL2 sources being  associated with  2/3/4/5/6 2RXS sources, respectively. The 
distribution of the positional uncertainties as a function of the flux in the detection band, colour 
coded by the likelihood of the detection, is presented in the right panel of Fig.~\ref{fig:2RXSproperties}.
Note, that figure shows the original positional uncertainty augmented by 5$\as$ in 
quadrature to account for  the systematic uncertainty on attitude reconstruction. The flux distribution 
(converted into the 0.5-2 keV band) shown in Fig.~\ref{fig:flux_distribution}. 

%--------------------------
%------------------------------------

\subsection{AllWISE catalog}
\label{sec:allwise}

The Wide-field Infrared Survey Explorer \citep[WISE\footnote{see also \url{http://irsa.ipac.caltech.edu/
Missions/wise.html} for a summary  and details and on the reactivated mission};][]{Wright:2010yq}, was 
launched in 2009 and over the course of one year scanned the entire sky at least twice in the 3.4 and 
4.6\,$\mu$m bands (hereafter W1, W2, respectively)  and at least once  in the 12 and 22\,$\mu$m bands 
(W3, W4). In the AllWISE data release\footnote{Available at \url{http://wise2.ipac.caltech.edu/docs/
release/allwise/}} \citep[November 13, 2013][]{Cutri:2013aa} all the available  data are combined, 
reaching a 5$\sigma$ limiting W1, W2, W3, and W4 magnitudes of better than 17.6, 16.1, 11.5, and 7.9 
(all in the Vega system) over 95\,\%  of the extragalactic sky ($|b| > $15$^\circ$). The coverage is 
inhomogeneous, being deepest at the Ecliptic Poles.

We generated two independent catalogues that include all AllWISE sources located within a radius of 120$
\as$ from an X-ray position listed in the 2RXS and  XMMSL2 catalogues, respectively. From each 
catalogues duplicated sources were removed. No additional filtering was performed.
This procedure results in two independent catalogues  of 6,252,516 unique entries for 2RXS and 1,288,533 
for XMMSL2, covering total unique areas of 368.81 deg$^2$  and 60.79 deg$^2$, respectively.

%------------------------------------

\section{NWAY in a nutshell}
\label{sec:application}

\nway has been developed for identifying the multi-wavelength counterparts to X-ray sources to multiple 
catalogues in a multi-dimensional parameter space (e.g., position  and positional uncertainty, density of 
sources, magnitudes, colours, variability, morphology, etc.) in a Bayesian framework. The code builds 
on the original work of \citet{Budavari:2008qy} who developed the algorithm for  simultaneously matching multiple 
catalogues  and enhances it by allowing sources to be present only in a subset of the 
catalogues  \citep[e.g.,][]{Pineau:2017qy}. Additionally,  \nway can either generate an internal prior 
for each source parameter following the implementation of the Maximum Likelihood Ratio as presented in e.g., 
\citet{Brusa:2007fp}, or use an external, pre-constructed prior. 

\nway has already been successfully applied in a number of studies, e.g., the identification of 
counterparts to ROSAT \citep[1RXS;][]{Voges:1999fk, Voges:2000qy} sources in the pilot SDSS-III/SEQUELS 
program \citep[][]{Dwelly:2017qy} using two optical bands, simultaneously; the 
search for counterparts to Chandra and XMM detections in the Extended Chandra Deep Field South 
\citep{Hsu:2014fj} using three independent catalogues (optical, near-infrared and 3.6 $\mu$m)  
simultaneously and with internally constructed priors (see for all the options the \nway manual at \url{https://github.com/JohannesBuchner/nway/raw/master/doc/nway-manual.pdf}). It has 
also been applied to 1RXS  and earlier XMM-Slew Survey (release 1.6, 26th Feb 2014) data on the BOSS 
imaging footprint \citep{Dwelly:2017qy}, adopting an external, mid-infrared based colour-magnitude prior, 
similar to the one  chosen in this work.

A comprehensive description of \nway is given in Appendix~\ref{A:nway} together with a verification  
using internally generated priors for XMM-COSMOS (see Appendix~\ref{A:TestCosmos}). In the following we 
focus on the application of the code to the scientific aim of the paper, the AllWISE counterparts to 
2RXS and XMMSL2.

The \nway code answers the question: {\it Considering the astrometric information (i.e. distance from 
the X-ray source, positional uncertainties, and number densities) and priors (e.g. magnitude and colour 
distribution), what is the posterior probability for each AllWISE source within a given radius from a 
2RXS or XMMSL2 detection to be the correct counterpart to the X-ray source?}. For the analytical 
details the reader is referred to Appendix~\ref{sub:math-probability}. In short, \nway first computes 
for each source in the AllWISE catalogue the Bayes factor considering only distance from 
the X-ray source, positional uncertainties, and number densities. Next, 
the Bayes factor is weighted by the mid-infrared magnitude-colour information (see \S~\ref{sec:prior}). 
Then, each AllWISE source is associated with the probability {\code p\_i} of being the right counterpart 
to a specific X-ray detection. In addition, {\it for each X-ray detection}, \nway provides the 
probability,  {\code p\_any},  that any of the AllWISE sources is
 the right counterpart. The higher  the 
value of {\code p\_any} the lower is the probability of a chance association.
In the output catalogue of \nway, for a given X-ray source, all the AllWISE within the search 
area are listed, ranked in decreasing order by their {\code p\_i}. For comfort \nway flags the first 
AllWISE source of each group as {\code match\_flag}=1, this being the best counterpart among those
available. A {\code match\_flag=2} indicates the AllWISE sources with a ${\code p\_i}/{\code p\_i}
_{best}<\alpha$ from the first, $\alpha$ being defined by the user (in this paper it is fixed to 0.5); 
these are considered secondary possible counterparts. Everything else is flagged as {\code match\_flag}=0. 
%------------------------------------
 
\section{Application of NWAY to 2RXS and XMMSL2}
\label{sec:prior}

In this section we motivate the AllWISE colour-magnitude prior, subsequently present the results of the 
application of \nway to the 2RXS and XMMSL2 catalogue defined in \S~\ref{sec:datasets}, and finish with 
the comparison of the associations for sources that are in common to both X-ray catalogues. 
%------------------------------------
\subsection{AllWISE colour-magnitude prior}
\label{sec:prior1}

The prior is defined as the probability, given
observable information alone, i.e. before considering any positional
information, that a counterpart is related to an X-ray source.
Given that the X-ray point-source population is an ensemble made of stars, nearby galaxies, and galaxies 
at unknown redshift hosting an AGN of unknown power, a prior based on a single magnitude distribution is 
insufficient. This is especially true for X-ray detections with large positional uncertainties. Ideally, 
the prior would use the entire SED as discriminator \citep[e.g.][]{Roseboom:2009kx}. In practice,  the 
lack of sufficiently deep multi-wavelength coverage of the entire sky requires a compromise. 

The AllWISE catalogue provides photometric coverage of the entire sky in the mid-infrared, a regime 
where the number density of sources is low compared with e.g., the optical bands. At the same time, 
virtually all point-like X-ray sources found in 2RXS and XMMSL2 are expected to be detected at the depth 
of the AllWISE survey, as we show later in this section.

To generate the prior we need to start with an X-ray sample that matches the sources expected at the 
depths of 2RXS and XMMSL2 but with secure counterpart association. Beyond a comparable flux limit this 
sample also needs to cover a sufficiently large area to include rare and bright objects. 
Both characteristics are fulfilled by the 3XMM-DR5 \citep{Rosen:2016kl} with a sky coverage of 
877\,deg$^{2}$  and with a flux limit  significantly deeper than 2RXS and XMMSL2. 
Following the same screening procedure outlined in \citet{Dwelly:2017qy}, but on the entire footprint of the survey we retained 2,349 sources 
distributed as in Fig.~\ref{fig:flux_distribution}.
All the sources selected in this way have a unique AllWISE counterpart 
within 5$\as$, 98\% within 3$\as$. Given the PSF of WISE, this provides a high confidence that the counterpart association is reliable. 

\begin{figure}
\centering
\includegraphics[width=8.0cm]{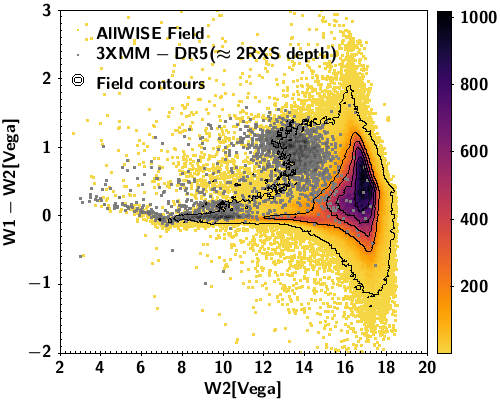}
     \caption{AllWISE colour-magnitude ( [W2] vs [W1-W2]) distribution of counterparts to the 3XMM-DR5 
catalogue cut at the depth of 2RXS (grey) compared with the AllWISE distribution (contours and density map) of 
all sources within 2$\am$ of the 3XMM-DR5 sources. \label{fig:wiseplots}
     }
\end{figure}

\begin{figure}
\centering
\includegraphics[width=8.0cm]{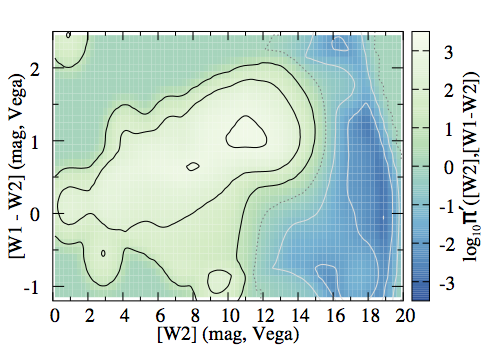}
\caption{Map of the weighting function, $\pi$, constructed from Fig.~\ref{fig:wiseplots} following the 
description in the text. Contours are drawn at $\log_{10}(\pi([W2],[W1-W2]) $= 3,2,1,0,-1,-2,-3. More 
description in \S~\ref{sec:prior1}.
\label{fig:prior1}
}
\end{figure}

The colour-magnitude distribution of the AllWISE counterparts to the 3XMM-DR5 sources is shown in 
Fig.~\ref{fig:wiseplots} together with the properties of the AllWISE field population. The 3XMM-DR5 
counterparts are well separated from the bulk of the AllWISE population, suggesting this colour-magnitude 
distribution to be an efficient prior. As in \citet{Dwelly:2017qy}, we generated a grid on the 
[W2],[W1-W2] plane with steps of 0.25\,mag in [W2] and 0.1\,mag in [W1-W2] (see Figure~\ref{fig:prior1}) and for 
each bin computed the ratio of the densities of 3XMM-DR5 counterparts and field sources. 
This two-dimensional distribution of density ratios encodes the prior
which we apply to the Bayes factor. The Bayes factor was computed taking into account astrometry (i.e. separation between the sources and respective positional uncertainty) and number density of the sources.

%------------------------------------
 
% ---------------------------------
 
% ---------------------------------
% ---------------------------------
\subsection{2RXS and AllWISE association}
\label{subsec:2RXS_ALLWISE}   

\begin{figure}
\centering
\includegraphics[width=8.0cm]{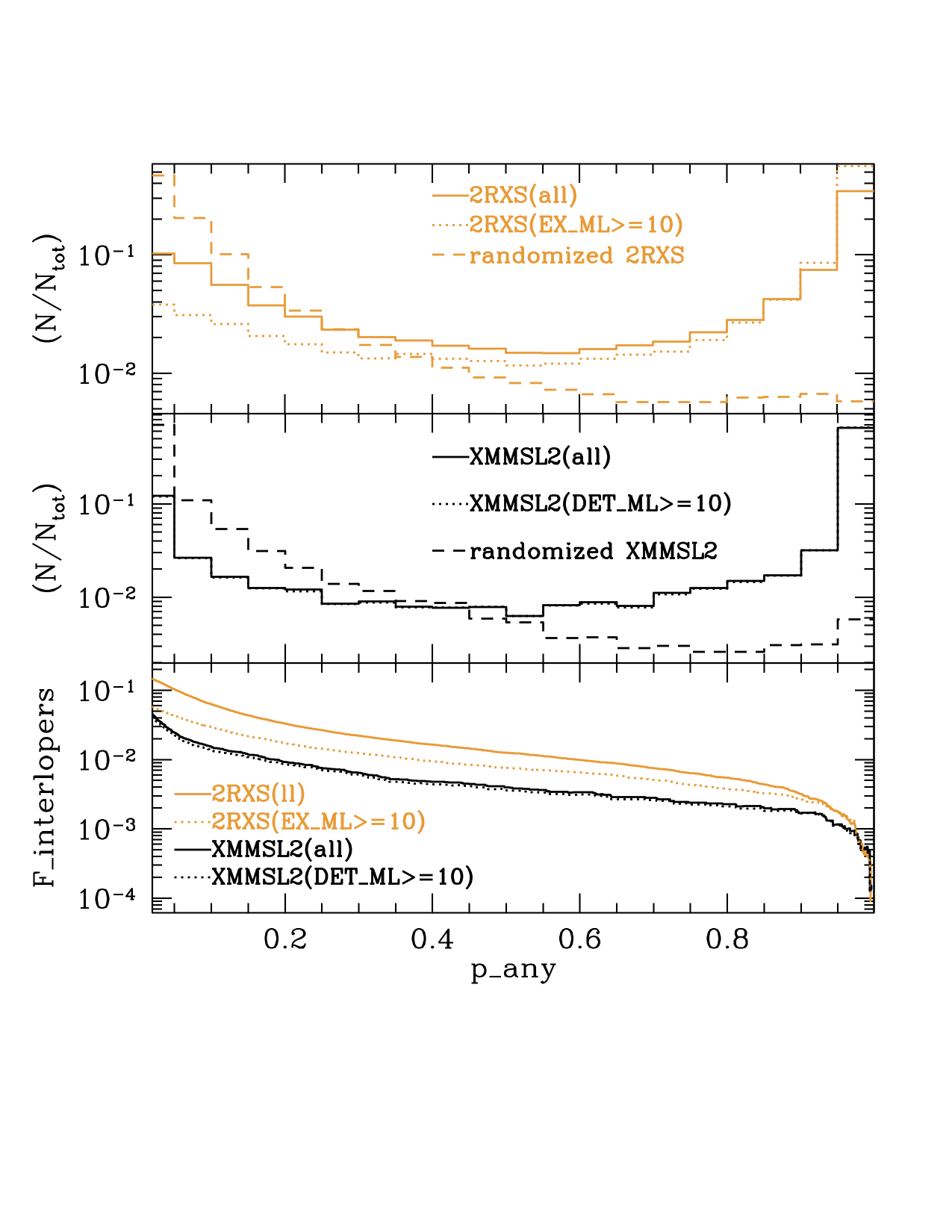}
\vspace*{-2.0cm}
\caption{Histogram distribution of the probability {\code p\_any} that the right counterpart is among the AllWISE sources for the 2RXS (top panel, gold) and XMMSL2 (middle panel, black) sources. The histogram is shown for the X-ray sources at the actual X-ray position (solid line) and after the systematic offset of the X-ray position (dashed line). The dotted lines show the distribution considering only the X-ray sources at the right position, with detection likelihood higher or equal 10. The similarity of the distribution in the case of XMMSL2 is justified by the high threshold of detection 
likelihood adopted in the original catalogue. The bottom panel shows at any given {\code p\_any} the 
fraction of interlopers, measured as the fraction of sources with {\code p\_any}$_{random} $>$ {\code p
\_any}_{real}$, for the complete samples and for the samples limited at the respective detection 
likelihood$\geq$10.}
\label{fig:posts_maps}
\end{figure}

We applied \nway  and the prior discussed in \S~\ref{sec:prior1} to $\approx6$\, Million AllWISE 
(see \S~\ref{sec:allwise}) sources within 2$\am$ from the 106,573 2RXS sources (see \S~ \ref{subsec:2rxs}).
At least one AllWISE candidate counterpart is found for all but 93 (0.01\%) of the 2RXS sources. A histogram of the distribution of  
{\code p\_any} is shown as the yellow solid line in the top panel of Fig.~\ref{fig:posts_maps}. 

\begin{figure}
\centering
\includegraphics[width=8.0cm]{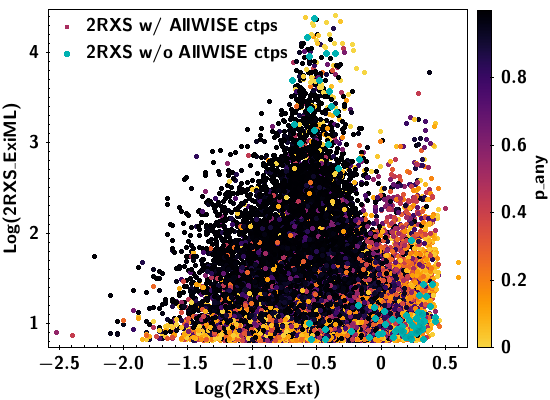}
\caption{X-ray Extension vs. detection likelihood for the 2RXS sources, colour coded as a function of 
{\code p\_any}.
Whilst sources with high {\code p\_any} are widely distributed, the sources with low {\code p\_any} are 
confined at low detection likelihood or significant extension. Green dots represent the sources for 
which a AllWISE counterpart was not found (see \S~ \ref{subsec:2RXS_ALLWISE} for more details).
\label{fig:exi_ext}}
\end{figure}

The 93 sources without any AllWISE counterpart (green points in Fig.~\ref{fig:exi_ext}) include a) extended X-ray sources (e.g. 2RXS\_J152238.4+083422, a spectroscopically confirmed cluster at z$\approx$0.035); b) X-ray sources with candidate counterparts present in the AllWISE images but not contained in the AllWISE catalog; and c) X-ray sources with very bright optical candidate counterparts not present in the AllWISE catalog.

63,305 2RXS sources ($\approx$59\% of the sample) have a {\code p\_any}$>0.5$ while for 35,571 sources ($
\approx$33\% of the sample) {\code p\_any} is lower than 0.3. Interestingly, 60\% of the latter are fainter than 14.5\,mag in W2. In this region the magnitude distribution of the prior  overlaps with  the bulk of the field population, indicating that the limit of the disentangling power 
of the prior has been reached and that the selected AllWISE counterpart could be the result of a chance association. This is partly due to astrometric scatter.

Next we investigate the typical {\code p\_any} for unreliable associations. We run  
\nway in the same configuration after a) shifting the coordinates of the 2RXS catalogue by 6$\am$ 
in Declination, b) extracting the AllWISE sources within 2$\am$ from the new 2RXS positions and c) 
removing the 2,059  sources (2\% of the sample)  that after the shift entered in the 2$\am$ radius 
circles from actual 2RXS sources. As expected, the distribution of {\code p\_any} (Figure~\ref{fig:posts_maps} top; yellow long-dashed line) peaks toward  low values of {\code p\_any}, with 78\,\% of the sample having {\code p\_any}$<\sim$0.15. This coincides with the idea that in any random sky position very few sources have properties matching the prior. E.g., only 5\,\% in the randomized 2RXS sample have {\code p\_any}$>$0.5 and {\code p\_i}$>$0.8. We can easily imagine that some of these sources are counterparts to actual X-ray sources below the 2RXS flux sensitivity. This hypothesis will be validated with future deeper X-ray data, e.g., from SRG/eROSITA. 

A very conservative {\code p\_any} $>$ 0.5 for a reliable association (thus with 
$<2$\,\% probability of chance association; see Figure~\ref{fig:posts_maps}) results in a sample of 
62,944 AllWISE counterparts to 2RXS sources. However, we release here the  entire catalogue of 2RXS 
counterparts, leaving to the user to decide the acceptable level of completeness and purity for their 
application. Figure~\ref{fig:posts_maps} (bottom) shows the fraction of expected interlopers 
for any given value of {\code p\_any}.

If we consider only sources with X-ray detection likelihood \citep[\code{EXI\_ML}; as defined in][]
{Boller:2016qy} larger than 10, the fraction of sources with {\code p\_any}$>$0.5 increases to 80\,\% 
(40,207/50,544). This means that many of the sources with low \code{p\_any} are among those with low 
detection likelihood, indicating that they could be just spurious detections. Alternatively, they could be real sources with poorer positional determination. The distribution of {\code 
p\_any} for the sources with \code{EXI\_ML}$>$10 is shown with the dotted line in Fig.~\ref{fig:posts_maps}.

\subsubsection{Multiple associations}

\begin{figure*}
  \centering
\includegraphics[width=8.0cm]{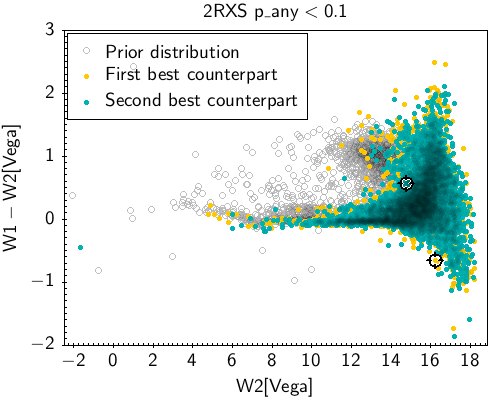}\includegraphics[width=8.0cm]{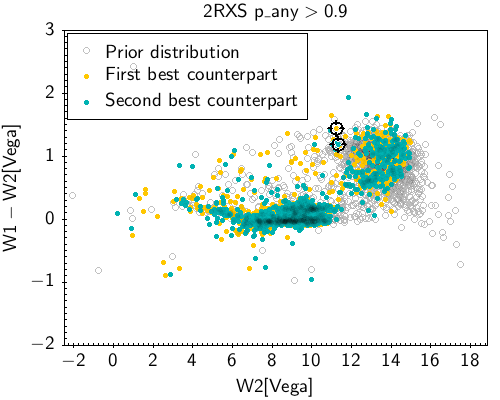}
\caption{Primary (gold) and secondary (green) possible AllWISE counterparts to the 5,844 and 899 
2RXS sources having two possible counterparts and {\code p\_any}$<$0.1 (left panel) and
{\code p\_any}$>$0.9 (right panel), respectively. The grey open circles represent the 3XMM-Bright sources used to build the prior. Two examples with {\code p\_any}$<$0.1 (2RXS\_J054219.4-080745) and {\code p\_any}$>$0.9 (2RXS\_J175642.5+512108) are highlighted in the left and right panel, respectively. A very similar result was obtained for XMMSL2 (see text) but is not shown here for simplicity. \label{fig:matchflag2}}

\end{figure*}

There are 17,734 (16.6\,\%) 2RXS sources with more than one candidate AllWISE 
counterpart\footnote{12321/3681/1177/386/121/34/11/2/1 cases with 2/3/4/5/6/7/8/9/10 AllWISE 
counterparts within the search area, respectively. Not only do most of these sources have a low 
\code{p\_any}, but the candidate counterparts are also faint in W2 and on average at larger distance from the X-ray position. This suggests that the X-ray sources themselves could be spurious. In fact, $>75$\,\% of this subsample have EXI\_ML$<10$. 
Only 7\,\% of them have {\code p\_any}$>0.9.$}, with the possible counterparts located in areas well populated by the prior. Given the poor angular resolution of ROSAT, it would not be a surprise if the candidate counterparts belong to distinct X-ray sources, detected as one in 
2RXS.
This is demonstrated in Figure~\ref{fig:matchflag2} (top), which shows the colour-magnitude 
distribution of the AllWISE sources for the 47\,\% (7\,\%) of the 12,321 2RXS sources with two candidate
counterparts having {\code p\_any}$<$0.1($>$0.9).

% -------------------------------
% -------------------------------

% -------------------------------
% -------------------------------
     
\begin{table}
\caption{XMMSL2 vs 2RXS  AllWISE association for sources in common }
\begin{center}
\begin{tabular}{l|c|c|c}
\hline
\multicolumn{2}{c}{XMMSL2-2RXS} & Sources in & Identical Best\\
Sep. & Mean Sep.& common &AllWISE ctp.\\
arcsec  & arcsec &    N    &   \% \\
\hline
\hline
$\le$5 &  3.2& 1145   & 98.5  \\
$\le$10 & 6.1& 3559 &  98.5   \\
$\le$30 & 12.4&  8202    &  95.7 \\
$\le$60 & 15.9&  9330  & 91.6  \\
\end{tabular}
\end{center}
\label{tab:XMMSL_2RXS}     
\end{table}     
% -------------------------------
% -------------------------------

% -------------------------------
\subsection{XMMSL2 and AllWISE association}
\label{subsec:XMMSL_ALLWISE}

The analysis done in the previous section was repeated for the XMMSL2-AllWISE association, with the 
summarizing plot being  in the middle and bottom panels of Fig.~\ref{fig:posts_maps}. 
First of all, the smaller X-ray positional error of {\it XMM} translates into a distribution of 
{\code p\_any} toward higher values (compare the solid and dashed cumulative curves in the right panel of the 
figure), with about 76\% of the sources having {\code p\_any}$>$0.5 and {\code p\_i}$>$0.8. Only 21\% of 
the sources have {\code p\_any}$<$0.3 with only 8 XMMSL2 sources without any AllWISE candidate 
counterpart within 2$\am$.
 
As for 2RXS, we systematically offset the positions of the XMMSL2 catalogue and run \nway with the same setting. 
Now we find that for only 3\% of the cases (571/17665), {\code p\_any}$>$0.5 and {\code p\_i}$>$0.8. The 
smaller positional  uncertainty also reduces the fraction of sources with more than one possible 
counterpart. In total 1210 XMMSL2 sources (6.8\%) have more than one possible counterpart\footnote{1015/163/25/17/1/1 sources having 2/3/4/5/6/7 possible counterparts. As for 2RXS, we analyze  
the properties of the XMMSL2 sample with two possible counterparts. Of the 1015 sources belonging to 
this group, 108 (10\%) have {\code p\_any}$>$0.9  and 739 (73\%) have {\code p\_any}$<$0.1}. As for 2RXS, 
also for XMMSL2 the majority of the sources with multiple associations have low {\code p\_any}, low 
magnitude distribution for the possible counterpart and, above all, low X-ray detection likelihood EXI\_ML\_B8$<10$.
Like for 2RXS, we will  provide all the associations in the catalogue, leaving the user to decide on the threshold for 
the reliability.

\subsection{2RXS vs. XMMSL2 associations}
\label{subsec:2RXS_XMMSL}

It is interesting to compare the association found for the sources that are in common to the 2RXS and XMMSL2 catalogues
as the smaller positional uncertainties of latter can give insight on the reliability of the 
associations for the former. 

Table~\ref{tab:XMMSL_2RXS} summarizes our results for the AllWISE counterparts that are 
in common for X-ray sources with matching coordinates within 5/10/30/60$\as$. Overall, the agreement between the associations is very good, with the highest fraction of identical counterparts found for the subset of X-ray sources with the smaller separation between 2RXS and XMMSL2.

\section{Comparison with literature}
\label{sec:literature} 

Since the release of the first ROSAT all-sky catalogues \citep{Voges:1999fk, Voges:2000qy} there have been many 
attempts to determine the multi-wavelength counterparts. Most of the follow-up of 
ROSAT point-like sources concentrated on the bright sources \citep[][]{Rutledge:2000dz, Schwope:2000ys, 
Mahony:2010oq}, even if, the adopted methodologies (e.g., association technique, secondary 
catalogues for the follow-up) changed over time. A direct comparison between those earlier works and the results presented here would only allow comparing the associations without assessing their correctness. In addition, the X-ray positions have changed from 1RXS to 2RXS (see \citealt{Boller:2016qy}) for more details. 

It is for this reason that we decided to test our associations against an astrometric reference sample of 4,524 X-ray 
sources from XMM, {\it Chandra} and {\it Swift} in the BOSS footprint, that have reliable counterparts \citep[see][for 
details on the sample]{Dwelly:2017qy}. A match on the X-ray positions within 60$\as$ provides 1,496 unique identifications in 
2RXS, while additional 14 2RXS sources have a second possible match. Of these, 1,418 have identical AllWISE 
counterparts corresponding to an accuracy of 94.8\%, with $
\approx$94\% of the identical associations having {\code p\_any}$>$0.5.

The exercise repeated for XMMSL2, results in identical AllWISE counterparts for 533 of the 547 sources 
that have a match within 30$\as$ in the reference catalog, corresponding to 97.4\% agreement. 514/533  
(96.4\,\%) have {\code p\_any}$>$0.5. This attests both the appropriateness of the prior and the reliability of \nway. \\
 
% -------------------------------

\section{Source Characterization}
\label{sec:characterization}

For the counterparts of the 2RXS and XMMSL2 all-sky surveys no single survey provides both photometry and spectroscopy over the full sky.
However, we can make an educated guess of the type of population by  a) matching with {\it Gaia}\footnote{\url{http://archives.esac.esa.int/gaia}} \citep{Arenou:2017kx}, b) studying the AllWISE colour distribution of the counterparts and comparing 
them with literature \citep[e.g.,][]{Wright:2010yq, Nikutta:2014ss}, and, finally 3) comparing Infrared 
and X-ray properties of the counterparts with those  well studied in the COSMOS field \citep[XMM-COSMOS, 
{\it Chandra}-COSMOS, {\it Legacy Chandra}-COSMOS;][respectively]{Brusa:2010lr,Civano:2012ys,Marchesi:2016jw}.

\subsection{2RXS and XMMSL2 counterparts in {\it Gaia}}
\label{subsec:GAIA}

The release of the first {\it Gaia} DR1 catalogue enables us to further characterize the  AllWISE 
counterparts of 2RXS and XMMSL2. In particular, it allows the identification of the sources with proper 
motion, indicating their Galactic nature. For this purpose we used the Hot Stuff for One Year (HSOY) catalogue \citep{Altmann:2017yq}.
HSOY includes 583,001,653 objects with precise astrometry based on the cross-match between the Catalogue of Positions and Proper 
Motions \citep[PPMXL;][]{Roeser:2010ys} and {\it Gaia} DR1 \citep{Arenou:2017kx}.
We find a HSOY match within 3$\as$ for 91427/132216 (70\%) and 14558/19120 (76\%) of all the AllWISE 
candidate counterparts (i.e. {\code match\_flag}=1 and {\code match\_flag}=2) to 2RXS and XMMSL2, respectively. 
Limiting the search only to the best AllWISE counterparts (i.e. {\code match\_flag}=1), we obtained a 
match with {\it Gaia} for  80078/106573 (75\%) and 14008/17665 (80\%). Of these, 10472/80078 (13\%)  and 
2054/14008 (15\%) have a measured proper motion (above 5$\sigma$) in the HSOY catalog, identifying them as Galactic objects.
 
%----------------------------

\subsection{IR/X-ray properties comparison with COSMOS}
\label{sec:COSMOS-2RXS}
  
\begin{figure}
  \centering
\includegraphics[width=8.0cm]{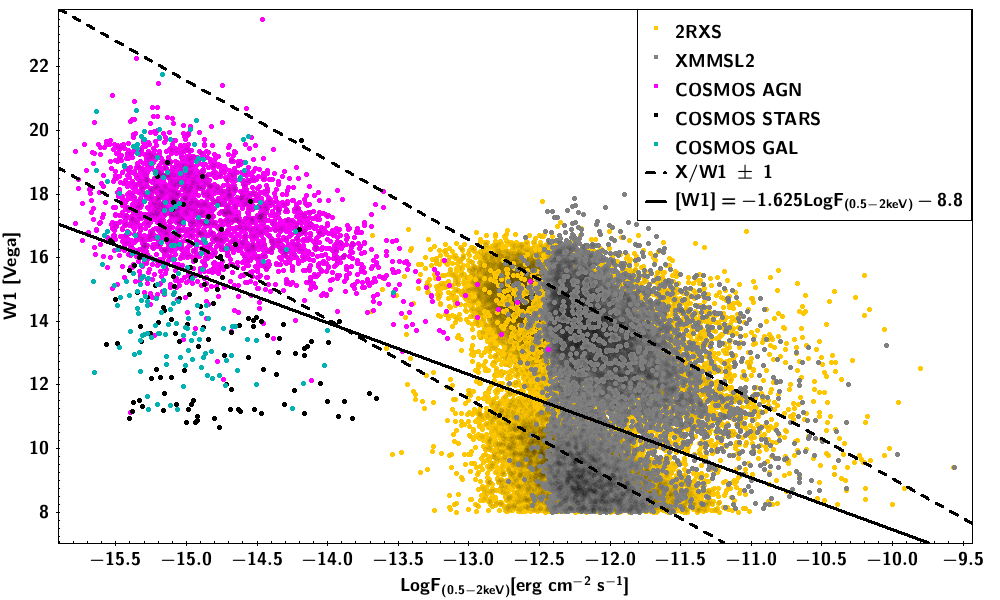}
\includegraphics[width=8.0cm]{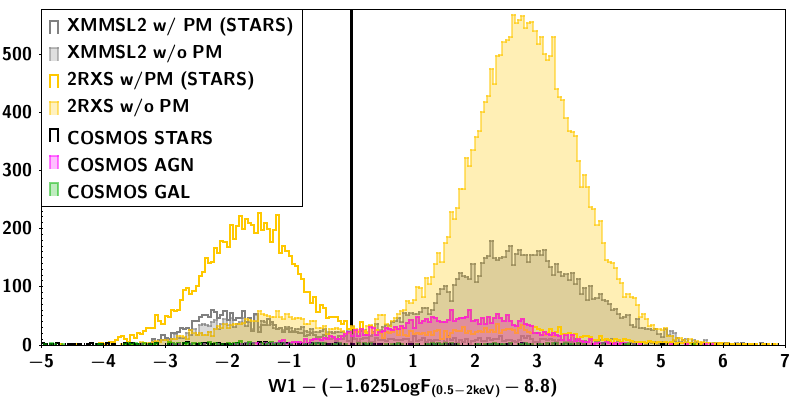}

\caption{{\bf Top:} W1 magnitude plotted against the 0.5-2 keV flux for the counterparts to {\it 
Chandra} Legacy-COSMOS survey (magenta=AGN, green=galaxies, black=stars) and for the AllWISE counterparts to 2RXS (yellow) 
and XMMSL2 (grey) sources with detection likelihood larger that 10 and {\code p\_any}$>$0.5. The dashed lines define the AGN locus 
historically defined by \citet{Maccacaro:1988lr}  and revised by \citet{Civano:2012ys} as described in 
\S~\ref{sec:COSMOS-2RXS}. The solid line has the slope as defined in Eq.~\ref{eq:slope} and best 
separated the star/non-star bimodal distribution of the sources in the three surveys. The cuts at [W1]$
\approx$11 and [W1]=8 correspond to the saturation limits for IRAC/[3.6] $\mu$m in COSMOS and [W1] in 
AllWISE. {\bf Bottom:} Histogram distribution of [W1] with respect to the solid line. Most of the 
sources below the line (left in this plot) are stars with a measured proper motion. Most of the sources 
above the line are supposed to be AGN as the distribution of the AGN in COSMOS would suggest.
\label{fig:2RXSCOSMOS}}
\end{figure}

Originally, \citet{Maccacaro:1988lr} noted that AGN found in the Einstein Observatory Extended Medium 
Sensitivity Survey \citep[EMSS;][]{Gioia:1987ys} were characterized by 
 log(f$_{\rm x}/{\rm f}_{\rm V})=\pm$1, with M stars and galaxies only marginally overlapping in this region. 
Since then, this locus was adapted by practically all X-ray surveys, extending the 
relation to other wavelengths ({\it r, i, K}, IRAC/[3.6\,$\mu$m]) and X-ray energy bands. The validity of  
the locus was confirmed with recent works \citep[e.g.,][]{Brusa:2007fp, Brusa:2010lr, 
Civano:2012ys}  pointing out that the near-infrared (e.g., {\it K}) or MIR (e.g., 3.6$\mu$m) 
bands provide a tighter correlation with X-rays than the optical bands. Here, however, the faintest of the X-ray AGN would 
fall below the locus (e.g., dashed line in Fig.~\ref{fig:2RXSCOSMOS}) and thus would overlap more with 
galaxies and stars.

In this paper we extend the earlier studies by combining the {\it Chandra} Legacy-COSMOS survey with 2RXS and 
XMMSL2. The {\it Chandra} Legacy-COSMOS  survey \citep{Civano:2016kb, Marchesi:2016jw} is preferred as it 
has a homogenous depth and covers sufficient area to host also some bright and rare sources. In addition,  
the counterparts are secure and well understood, thanks to the depth and amount of the available ancillary data. Furthermore, the spectroscopic follow-up and the reliable photometric redshifts via SED fitting \citep{Marchesi:2016jw, Salvato:2011mz} allow the classification of the sources as Type1 
(unobscured) and Type2 (obscured) AGN, Galaxies (sources with ${\rm L_{X}<10^{42} erg/s}$), and stars. 
Fig.~\ref{fig:2RXSCOSMOS} (top) compares the properties of the 
counterparts in COSMOS, 2RXS, and XMMSL2 in the W1 vs. X-ray flux plane.
The AllWISE/W1 total magnitude for the {\it Chandra} Legacy-COSMOS sources has been derived from the 
flux in IRAC/[3.6\,$\mu$m]  within 1.9$\as$ aperture as listed in  \citet{Laigle:2016dq} using the 
conversion factor 0.765 and transforming AB to Vega magnitudes as prescribed by the S-COSMOS 
documentation available through the Infrared Science Archive 
\citep[IRAS\footnote{\url{http://irsa.ipac.caltech.edu/data/COSMOS/tables/scosmos/scosmos_irac_200706_colDescriptions.html}}; see also][]{Sanders:2007qd}.
The additional correction of W1 - [3.6] = 0.01\,mag, following \citet{Stern:2012kl} was 
applied.

In the same figure, when plotting the 2RXS and XMMSL2 sources, we considered for clarity only those with 
a detection likelihood  larger than 10, {\code p\_any}$>$0.5 and with an unique AllWISE counterpart.
In the figure, the dashed line correspond to the locus defined by \citep{Maccacaro:1988lr}, 

\begin{equation}
\label{eq:XW1}
X/O = log(f_{\rm X}/f_{\rm opt}) = log(f_{\rm X}) + C + m_{\rm opt}/2.5= \pm1 
\end{equation}

but using the flux at 0.5-2 keV band and the W1 magnitude, instead of 0.3-3.4\,keV and m$_{\rm opt}$=V-band, respectively. The coefficient C takes into account the different effective central wavelength and width of the filters.

The solid line is empirical and defined as

\begin{equation}
\label{eq:slope}
[W1] = -1.625*logF_{(0.5-2 keV)} - 8.8
\end{equation}

This new relation much better separates AGN from galaxies and stars over six orders of magnitude and passes 
through the bimodal distribution of the counterparts to 2RXS and XMMSL2. As for COSMOS, most of the 2RXS 
and XMMSL2 sources  below this relation are stars with a well measured proper motion as described in 
\S~\ref{subsec:GAIA}. A complementary way to visualize this natural separation is to plot the distribution of the sources with with respect to the solid line (Fig.~\ref{fig:2RXSCOSMOS} bottom). Here, stars are indicated with a solid line, while non-stars are represented 
with filled histograms. Interestingly, 98.7\% of the spectroscopically confirmed X-ray selected AGN presented 
in \citet{Dwelly:2017qy} lie above the solid line. 
Similarly, only 0.02\% of all the AllWISE counterparts to 2RXS classified as {\it stars} via their proper motion, lie above the relation.
We suggest to use this new empirical X-ray/MIR relation as a straightforward mean to separate stars and quasars in samples of point-like X-ray sources. 
In fact, we show in the next section that most of the sources below the relation, despite not having
measured proper motions, are also stars  based on their AllWISE colours. Inversely, only 0.03\% of the 
AllWISE counterparts to 2RXS and XMMSL2 that are classified as AGN using the WISE colours as defined by 
\citet{Stern:2012kl, Assef:2013sp}, lie below the solid line.

%---------------------------------

%--------------------- 
 
\subsection{IR properties of 2RXS and XMMSL2 counterparts}

\begin{figure*}
  \centering
\includegraphics[width=15cm]{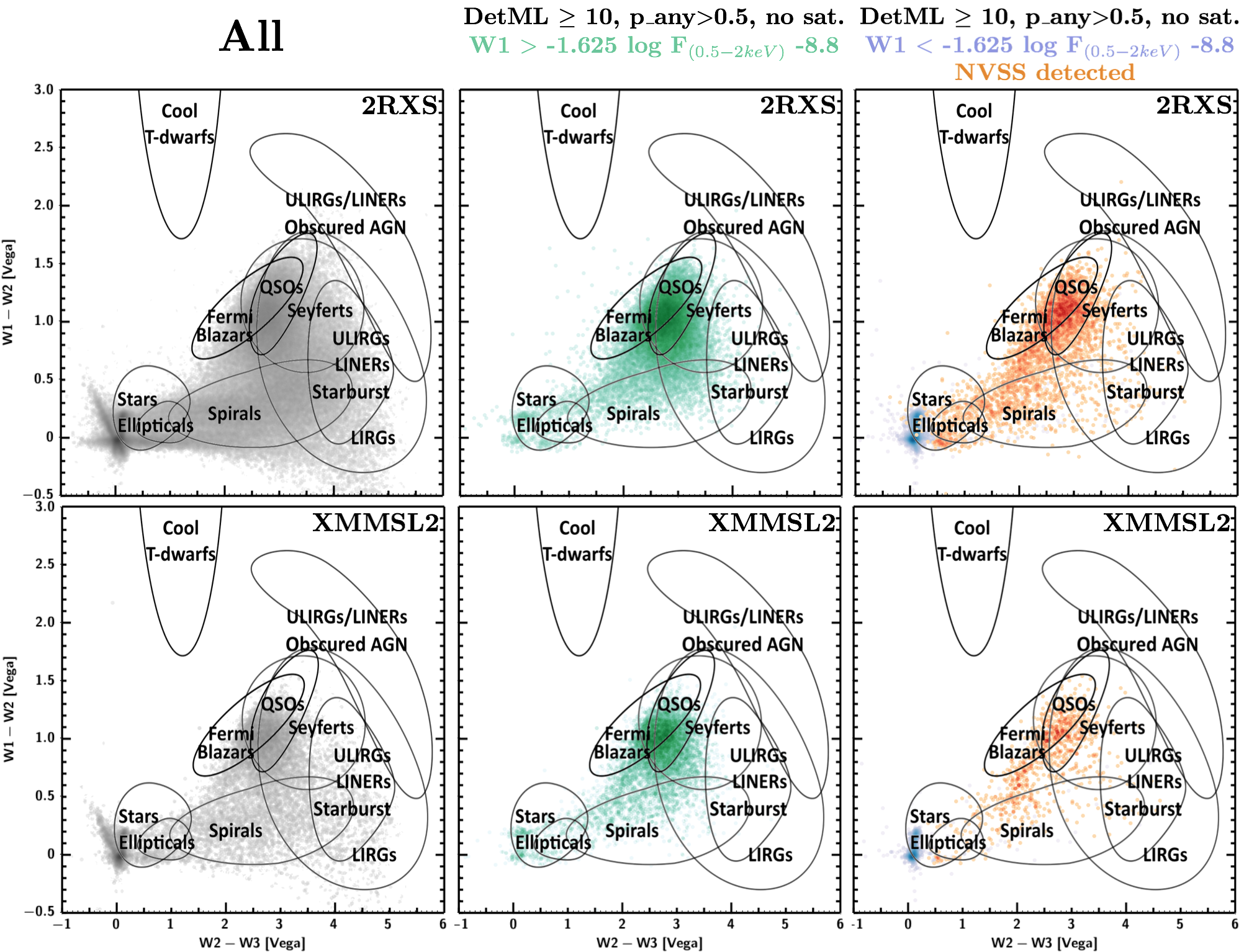}
\caption{Density distribution of the AllWISE of counterparts to 2RXS and XMMSL2 plotted over the colour-colour diagram originally created by Chao-Wei Tsai, (used here with 
permission) in \citet{Wright:2010yq}, but modified  by adding the approximative locus of the counterparts to Fermi sources 
\citep[e.g.,][]{DAbrusco:2013kn}. {\bf Top:} the AllWISE counterparts are plotted for all the 2RXS. The sources with higher detection likelihood, with more reliable and non saturated counterparts are plotted in the middle and right panels.
The sources are further split with respect to the relation defined in equation 2: the 25,000 green sources are above the relation, i.e. expected to be dominated by AGN); of these, $\approx$ 3,450 have a counterpart in the NVSS catalog. In bluish colour we plot the $\approx$ 9,500 sources that are below the relation and are expected to be mostly stars.
{\bf Bottom:} same as in the top, but for XMMSL2 sources. There are 7,259 sources dominated by AGN, 2,168 stars and 891 with a NVSS counterpart, respectively.
The properties of the sources correspond to the expected one based on their location in the AllWISE colour-colour plot.\label{fig:wright}}
\end{figure*}

The AllWISE colours [W1-W2] and [W2-W3]\footnote{0.02\% (0.08\%), 0.07\% (0.3\%) and 20\% (20\%) are only 
upper limits in W1,W2,W3 in 2RXS (XMMSL2), respectively.} of the candidate counterparts can be used their 
qualitative characterization, as suggested by \citet[e.g.][]{Wright:2010yq, Nikutta:2014ss}. 
Fig.~\ref{fig:wright} shows the AllWISE colours of the 2RXS (top) and XMMSL2 (bottom) counterparts, using in 
background Fig. 12 of \citet{Wright:2010yq}. To the well known loci we added the location of the 
Fermi/Blazars identified by, e.g. \citet[][]{DAbrusco:2013kn}. That is also the location of most of the  X-ray 
sources that are associated with radio emitters \citep[e.g., NVSS:][]{Condon:1998vn}; 4308 sources in 2RXS 
and 1307 in XMMSL2, respectively. As suggested by \citet{Tsai:2013qy}, the sources in this locus are 
nearby, jetted objects (z$<$0.5),  suggesting indeed the presence of an AGN in their cores (Emonts private 
communication, Emonts et al, in prep).
As expected, the bulk of the X-ray population in 2RXS and XMMSL2  is characterized by QSO, AGN, 
and stars. In particular, the sources below the relation \ref{eq:slope} are concentrated in the stellar locus, while the bulk of the sources with W1 above the relation are in the AGN/QSO loci.

% -------------------------------

% -------------------------------

\section{catalogues release}
\label{sec:catalogues}
We release the AllWISE associations to the sources in the 2RXS and XMMSL2 catalogues outside the 
Galactic plane.
The list and the description of columns are provided for each catalogue in the two following sections.
In short, we provide few columns that are keys to the identification of the X-ray sources, simply 
extracted, without any modification, from their original catalogues. We complement each source with the 
list of possible AllWISE counterparts and the output columns of \nway that are essential for those users 
interested in defining more pure and complete subsamples. We provide columns that inform the user on 
whether or not there is more than one possible counterpart. Finally, the data are complemented with a 
match to the {\it Gaia} DR1 catalog.
A simple match with the unique identifier from 2RXS, XMMSL2, AllWISE, 2MASS and {\it Gaia} will allow 
the user to retrieve additional columns from the original catalogues, not listed in our release.
The catalogues accompany  this paper but will be available also via Vizier and at the dedicated web page \url{http://
www.mpe.mpg.de/XraySurveys/2RXS_XMMSL2/}.
 
\subsection{2RXS-AllWISE catalogue}
\label{subsec:2RXS2ctps}

{\it Column 1.} {\bf 2RXS\_ID}: IAU Identifier from \citet{Boller:2016qy}.\\
{\it Columns 2-3.} {\bf 2RXS\_RA, 2RXS\_DEC}: 2RXS J2000 Right Ascension and Declination, in degrees.\\
{\it Column 4.} {\bf 2RXS\_e\_RADEC}: 2RXS positional error, in arc seconds.\\
{\it Column 5.} {\bf 2RXS\_ExiML}: 2RXS source Detection Likelihood. User should refer to the \citet{Boller:2016qy} for discussion on the fraction of false detections as function of this parameter.\\
{\it Column 6.} {\bf 2RXS\_Ext}: 2RXS source extent in units of image pixels.\\
{\it Column 7.} {\bf 2RXS\_ExtML}: Probability of the 2RXS source extend.\\
{\it Column 8.} {\bf 2RXS\_SRC\_FLUX}: 2RXS flux in unit of $\cgs$ \citep[see][for details]{Dwelly:2017qy}.\\
{\it Column 9.} {\bf 2RXS\_SRC\_FLUX\_ERR}: 2RXS flux error \citep[see][for details]{Dwelly:2017qy}.\\
{\it Column 10.} {\bf ALLW\_ID}: WISE All-Sky Release Catalogue name \citep{Cutri:2013aa}.\\
{\it Columns 11-12.} {\bf ALLW\_RA, ALLW\_DEC  }: J2000 AllWISE Right Ascension and Declination, in degrees.\\
{\it Column 13.} {\bf ALLW\_e\_RADEC}: AllWISE positional error, in arc seconds.\\
{\it Columns 14-17.} {\bf ALLW\_w[1234]mpro}: AllWISE Vega magnitude in the W1, W2, W3, W4 bands.\\
{\it Columns 18-21.} {\bf ALLW\_w[1234]sigmpro}: AllWISE magnitude error in the W1, W2, W3, W4 bands.\\
{\it Columns 22-25.} {\bf ALLW\_w[1234]snr}: AllWISE signal to noise ratio in the W1, W2, W3, W4 bands.\\
{\it Column 26.} {\bf ALLW\_cc\_flags}: AllWISE reliability flag from \citet{Cutri:2013aa}.\\
{\it Column 27.} {\bf Separation\_ALLW\_2RXS}: Separation between members of this association, in
arcsec.\\
{\it Column 28.} {\bf dist\_bayesfactor}: Logarithm of ratio between prior and posterior from distance 
matching.\\
{\it Column 29.} {\bf dist\_post}: Distance posterior probability comparing this association vs. no 
association, as in \citep{Budavari:2008qy}.\\
{\it Column 30.} {\bf bias\_ALLW\_COLOURMAG\_PIX}: Probability weighting introduced by AllWISE prior. 1 
indicates no change.\\
{\it Column 31.} {\bf p\_single}: Same as dist\_post, but weighted by the AllWISE color-magnitude prior.
\\
{\it Column 32.} {\bf p\_any}: For each entry in the X-ray catalogue, the probability that any of
the associations is the correct one. The lower {p\_any}, the lower is confidence that a reliable counterpart 
was found. See \S~\ref{subsec:2RXS_ALLWISE}.\\
{\it Column 33.} {\bf p\_i}: Relative probability of the match, if one exists. The {\code p\_i} add up to unity for each X-ray source.\\
{\it Column 34.} {\bf match\_flag}: 1 for the most probable match, if existing; 2: almost as good solutions {p\_i}/{p\_i}$_{\rm best} >$ 0.5).\\
{\it Column 35-36.} {\bf GroupID, GroupSize}: if the 2RXS source has only one possible AllWISE counterpart, the two columns are blank.
Otherwise, the GroupSize value indicate the number of possible counterparts while the GroupID value is 
the same integer for the group. A sort on the GroupID value, will rank the first non-unique match group 
together, followed by all the rows in the second non-unique group, etc. All the unique matches are 
listed last.\\
{\it Column 37.} {\bf 1RXS\_ID}: Source name in the 1RXS catalogues \citep{Voges:1999fk, Voges:2000qy}.\\
{\it Column 38.} {\bf ALLW\_2MASS\_ID}: 2MASS Identifier as listed in the AllWISE catalog.\\
{\it Columns 39-41.} {\bf ALLW\_[jhk]\_m\_2mass}: 2MASS magnitude in the j,h,k bands, as from AllWISE 
catalogue.\\
{\it Columns 42-44.} {\bf ALLW\_[jhk]\_msig\_2mass}: 2MASS magnitude errors in the j,h,k bands, as from 
AllWISE catalogue.\\
{\it Columns 45.} {\bf Gaia\_DR1\_ID}: Solution ID from the original {\it Gaia} DR1 catalogue 
\citep[see][for more details]{Fabricius:2016ul}.\\
{\it Columns 46-47.} {\bf Gaia\_DR1\_RA, Gaia\_DR1\_DEC}: {\it Gaia} J2000 Right Ascension and 
Declination as computed by Vizier.\\
{\it Columns 48-49.} {\bf pmra, pmdec}: Proper motion in Right Ascension and Declination as measured by 
{\it Gaia}.\\
{\it Columns 50-51.} {\bf pmra\_error, pmdec\_error}: Proper motion errors in Right Ascension and 
Declination as measured by {\it Gaia}.\\
{\it Columns 52.} {\bf phot\_g\_mean\_flux}: {\it Gaia} mean flux in units of e-/s.\\
{\it Columns 53.} {\bf phot\_g\_mean\_flux\_error}: {\it Gaia} mean flux error in units of e-/s.\\
{\it Columns 54.} {\bf phot\_g\_mean\_mag}: {\it Gaia} mean magnitude.\\
%----------------------------------

\subsection{XMMSL2-AllWISE catalogue}
\label{subsec:XMMSL2ctps}

{\it Column 1.} {\bf XMMSL2\_ID}: Unique identifier from \citet{Boller:2016qy}.\\
{\it Columns 2-3.} {\bf XMMSL2\_RA, XMMSL2\_DEC}: 2RXS J2000 Right Ascension and Declination, in degrees.\\
{\it Column 4.} {\bf XMMSL2\_e\_RADEC}: XMMSL2 original positional uncertainty augmented by 5$\as$ in 
quadrature.\\
{\it Columns 5-7.} {\bf XMMSL2\_DET\_ML\_B[876]}: XMMSL2 source Detection Likelihood in the respective 
energy bands.\\
{\it Column 8-10.} {\bf XMMSL2\_Ext\_B[876]}: XMMSL2 source extent in units of image pixels, in the 
respective energy bands.\\
{\it Column 11-13.} {\bf XMMSL2\_Ext\_ML\_B[876]}: Probability of the XMMSL2 source extend in the 
respective energy bands.\\
{\it Column 14-16.} {\bf XMMSL2\_FLUX\-B[876]}: XMMSL2 flux in the respective energy bands, in $\cgs$ 
units.\\
{\it Column 17-19.} {\bf 2RXS\_FLUX\_B[876]\_ERR}:  XMMSL2 flux errors in the respective energy bands, 
in $\cgs$ units.\\
{\it Column 20.} {\bf ALLW\_ID}: WISE All-Sky Release catalogue name \citep{Cutri:2013aa}.\\
{\it Columns 21-22.} {\bf ALLW\_RA, ALLW\_DEC  }: J2000 AllWISE Right Ascension and Declination, in degrees.\\
{\it Column 23.} {\bf ALLW\_e\_RADEC}: AllWISE positional error, in arc seconds.\\
{\it Columns 24-27.} {\bf ALLW\_w[1234]mpro}: AllWISE Vega magnitude in the W1, W2, W3, W4 bands.\\
{\it Columns 28-31.} {\bf ALLW\_w[1234]sigmpro}: AllWISE magnitude error in the W1, W2, W3, W4 bands.\\
{\it Columns 32-35.} {\bf ALLW\_w[1234]snr}: AllWISE signal to noise ratio in the W1, W2, W3, W4 bands.\\
{\it Column 36.} {\bf ALLW\_cc\_flags}: AllWISE reliability flag from \citet{Cutri:2013aa}.\\
{\it Column 37.} {\bf Separation\_ALLW\_XMMSL2}: Separation between members of this association, in
arcsec.\\
{\it Column 38.} {\bf dist\_bayesfactor}: Logarithm of ratio between prior and posterior from distance 
matching.\\
{\it Column 39.} {\bf dist\_post}: Distance posterior probability comparing this association vs. no 
association, as in \citep{Budavari:2008qy}.\\
{\it Column 40.} {\bf bias\_ALLW\_COLOURMAG\_PIX}: Probability weighting introduced by AllWISE prior. 1 
indicates no change.\\
{\it Column 41.} {\bf p\_single}: Same as dist\_post, but weighted by AllWISE prior.\\
{\it Column 42.} {\bf p\_any}: For each entry in the X-ray catalogue, the probability that any of
the associations is the correct one. The lower p\_any, lower is confidence that a reliable counterpart 
was found. See \S~\ref{subsec:2RXS_ALLWISE}.\\
{\it Column 43.} {\bf p\_i}: Relative probability of the match, if one exists. The {\code p\_i} add up to unity 
for each X-ray source.\\
{\it Column 44.} {\bf match\_flag}: 1 for the most probable match, if existing; 2: almost as good 
solutions (p\_i/ p\_i$_{best} $> 0.5).\\
{\it Column 45-46.} {\bf GroupID, GroupSize}: if the 2RXS source has only one possible AllWISE 
counterpart, the two columns are blank.
Otherwise, the GroupSize value indicate the number of possible counterparts while the GroupID value is 
the same integer for the group. A sort on the GroupID value, will rank the first non-unique match group 
together, followed by all the rows in the second non-unique group, etc. All the unique matches are 
listed last.\\
{\it Column 47.} {\bf 1RXS\_ID}: Source name in the 1RXS catalogues \citep{Voges:1999fk, Voges:2000qy}.\\
{\it Column 48.} {\bf ALLW\_2MASS\_ID}: 2MASS Identifier as listed in the AllWISE catalog.\\
{\it Columns 49-51.} {\bf ALLW\_[jhk]\_m\_2mass}: 2MASS magnitude in the j,h,k bands, as from AllWISE 
catalogue.\\
{\it Columns 52-54.} {\bf ALLW\_[jhk]\_msig\_2mass}: 2MASS magnitude errors in the j,h,k bands, as from 
AllWISE catalogue.\\
{\it Columns 55.} {\bf Gaia\_DR1\_ID}: Solution ID from the original {\it Gaia} DR1 catalogue 
\citep[see][for more details]{Fabricius:2016ul}.\\
{\it Columns 56-57.} {\bf Gaia\_DR1\_RA, Gaia\_DR1\_DEC}: {\it Gaia} J2000 Right Ascension and 
Declination as computed by Vizier.\\
{\it Columns 58-59.} {\bf pmra, pmdec}: Proper motion in Right Ascension and Declination as measured by 
{\it Gaia}.\\
{\it Columns 60-61.} {\bf pmra\_error, pmdec\_error}: Proper motion errors in Right Ascension and 
Declination as measured by {\it Gaia}.\\
{\it Columns 62.} {\bf phot\_g\_mean\_flux}: {\it Gaia} mean flux in units of e-/s.\\
{\it Columns 63.} {\bf phot\_g\_mean\_flux\_error}: {\it Gaia} mean flux erorr in units of e-/s.\\
{\it Columns 64.} {\bf phot\_g\_mean\_mag}: {\it Gaia} mean magnitude.\\
 
%------------------------------
\section{NWAY release}
\label{sec:nway}

Together with the AllWISE counterparts to the 2RXS  and XMMSL2 catalogues, we also also \nway.
The \nway software and manual are available at \protect\url{https://github.com/JohannesBuchner/nway}. In 
order to make the user familiar with the code, the release is completed with the catalogues used in the 
testing phase discussed in  Appendix~\ref{A:TestCosmos}. 
We would like to stress that the use of \nway is not limited to finding the counterparts to X-ray 
sources.
With the advent of deep and wide area surveys in  X-rays (e.g. eROSITA, Athena) and radio  (e.g., ASKAP/
EMU:\citealt{Norris:2011kl}; LOFAR:\citealt{van-Haarlem:2013tg}; APERTIF:\citealt{Oosterloo:2010kl}), \nway 
will provide a powerful and reliable counterpart identification tool. 

\section{Discussion and Conclusions}
\label{sec:discussion}

We presented the catalogues of secure AllWISE counterparts to the ROSAT/2RXS and XMMSL2 X-ray 
extragalactic all-sky surveys. Only a small fraction (less than 5\%) of the X-ray/AllWISE 
associations is expected to be due to chance associations. Associations were obtained using a new algorithm, \nway, capable of handling complicated priors. 
In particular, we have used here a prior based on the WISE colour-magnitude  properties of about 2500 X-ray
sources from the 3XMM-DR5 catalogue with flux distribution similar to 2RXS and XMMSL2.

\nway can be used for finding the right counterparts to other
(not only X-ray) surveys. However, the prior which we apply \nway in
this work is tuned to the properties of the input catalogs and thus is not universal. E.g., adopting  for 2RXS a similar prior to that used in \citet{Dwelly:2017qy}, which 
was constructed with half of the sources adopted here, the AllWISE counterpart changes for 3\% of the 
sources (3431/106573). The prior is appropriate only as long as it well represents the population.
For this reason, the prior adopted for the extragalactic region covered by 2RXS and XMMSL2 cannot be 
used with the same reliability for finding the correct counterparts of X-ray sources in the Galactic 
plane, where the X-ray catalogues are dominated by stars. Similarly, it will not be possible to use the same prior with the same 
reliability for finding the counterparts to X-ray surveys that are significantly shallower or deeper than the two discussed in this work.

\subsection{Finding counterparts to eROSITA point-like sources}
\label{subsec:eROSITA}

The design of \nway was dictated by the need of developing  a flexible algorithm that could 
be used with the patchwork of multi-wavelength coverage of the entire sky available for finding the 
counterparts of eROSITA \citep{Merloni:2012uq}.

eROSITA combines a wide field of view, large collecting area, long
survey duration, broad energy bandpass, and good point source location
accuracy, making it by far the most powerful X-ray survey instrument
ever built. In the soft energy band (0.5--2\,keV), the planned
four-year eROSITA all-sky survey, will have a median point source flux limit of $10^{-14}$\,$\cgs$\ \citep{Merloni:2012uq}, approximately 30$\times$ deeper than the ROSAT all-sky survey (for AGN-like X-ray spectra).

\begin{figure}[H]
  \centering
\includegraphics[width=8.0cm]{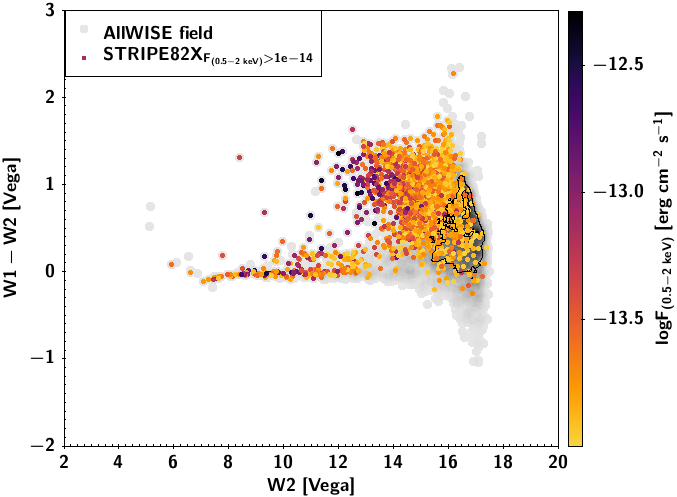} 
     \caption{WISE colour-magnitude plane for the AllWISE counterparts to the sources in STRIPE82X \citep{Tasnim-Ananna:2017gf}, cut at the expected depth of eROSITA at the end of the survey (eRASS:8). 
Sources are colour coded as a function of their X-ray flux. In grey and with black contours, the AllWISE population within 30\as~ from the STRIPE82X sources are shown. Given its shallowness AllWISE can provide a counterpart only to 80\% of the eROSITA sources. For the faint eROSITA sources, the AllWISE counterparts will increasily overlap with the bulk of the AllWISE population, thus reducing the disentangling power of the prior, even though the search for the correct counterpart is limited to 30\as}. \label{fig:eROSITA_WISE} 
\end{figure}

In the hard X-ray band
(2--10\,keV), the predicted flux limit of $2\times10^{-13}$\,\cgs\ is around 100$\times$ deeper than the 
only
existing all-sky survey conducted at these energies
\citep[i.e. the High Energy Astronomy Observatory, HEAO-I:][]{Wood:1984fu}. On completion, the eROSITA
survey is expected to detect about 4 millions X-ray sources, with 3/4 of them being AGN.
Thankfully, the location accuracy for point-like
eROSITA sources is expected to be better than 10$\as$ radius 
(combination of statistical and systematic uncertainties),
substantially better than for typical ROSAT 
sources. This will also be enabled by the availability of {\it Gaia} that will allow accurate positional 
accuracy on eROSITA single frame by tying the two astrometric reference frames. 
The eROSITA data will also enable better separation between 
point-like (mostly AGN and stars) and extended (galaxy cluster) sources 
on the basis of their X-ray properties alone \citep{Merloni:2012uq}.
However, due to the fainter X-ray flux limit
expected, the optical-IR counterparts to point-like eROSITA sources
will typically be several magnitudes fainter than those presented
here. Figure~\ref{fig:eROSITA_WISE} illustrates this by showing the colour-magnitude distribution of the counterparts to the X-ray sources in STRIPE82X \citep[][]{LaMassa:2016cq,Tasnim-Ananna:2017gf} cut at the depth of eROSITA, colour coded as a function of X-ray flux. 
Given the increasing depth of eROSITA, the counterparts of the sources get progressively fainter, 
finally overlapping with the bulk of the AllWISE population within 30$\as$ of the X-ray position, here in grey (see for comparison the 
distribution in Fig.~\ref{fig:wiseplots}). 

Hence, in order to select the correct counterparts for several
million eROSITA sources, we will need to take into account additional information to
separate field populations from the true counterparts to X-ray
sources. Deeper WISE catalogues, enabled by the co-addition of the
ongoing multi-year NEOWISE survey data
\citep{Mainzer:2011lh, Mainzer:2014mw, Meisner:2016bs} with the existing AllWISE data
set, should not only probe to fainter $[W1]$ and $[W2]$ limits, but should also have a smaller 
photometric
scatter at the magnitudes currently probed by the AllWISE survey. Such a
reduced scatter will allow better separation of the red (in $[W1$-$W2]$) AGN
population from the bluer field stars and galaxies. 
Note however that at the depth of ROSAT, only 0.01\% of the AllWISE counterparts had an upper limit in 
W2, while the number will increase to fainter X-ray flux, even considering the reactivation of NEOWISE 
\citep{Mainzer:2014mw, Meisner:2017ij} post-cryogenic phase expected to reach a depth in W2 of 19.9, 
when combined with WISE.

In addition, we expect
that one of the main drawbacks of relying on any catalogue derived
from WISE data will be the relatively broad PSF ($\sim$6\,arcsec
full width at half maximum in WISE bands 1 and 2), which
results in blending problems for close pairs of sources. This problem
will inevitably get worse as the co-added WISE data reach to
fainter magnitudes, approaching the confusion limit. 
In addition, once an AllWISE counterpart has been
selected for each X-ray source, the final step of optical counterpart
selection must still be carried out. This step becomes particularly
difficult when WISE detections are blends of multiple astrophysical
sources.

The forced photometry techniques and tools described by \citet{Lang:2016hc}
avoid many of the problems associated with combining data across
multiple wavebands, and have already been exploited successfully,
e.g. in the selection of QSO targets for eBOSS  \citep{Myers:2015ij}. By
cross-matching eROSITA sources with previously compiled forced
photometry catalogues (e.g. derived from  {\it Gaia} in the Galactic plane, SDSS and DECaLS, DES, VHS 
photometry), we expect to greatly reduce both the impact of source
confusion in the IR, and the general problems related to compiling
data across multiple optical-IR wavebands.

The high cross-matching success rate for 2RXS and XMMSL2  has
demonstrated that our cross-matching routine and priors are relatively
robust. However, the dynamic range of
the eROSITA catalogue will be much larger than that considered here.
  
Therefore, it is likely that a single,
X-ray flux-independent prior (as adopted in this work) will be a sub-optimal
choice for finding counterparts to \textit{all} eROSITA sources. We
also expect a strong dependence in the mixture of object classes
which make up the eROSITA sample as a function of Galactic latitude.
Thankfully, the XMM, Chandra and {\it Swift}/XRT archives already contain
large samples of well-measured X-ray reference sources which populate
the entire eROSITA  flux range, and which can be used to define new
X-ray-flux-dependent and/or Galactic-latitude-dependent optical-IR
priors.

However, great care will be needed to understand the very
complex inhomogeneities/biases/incompletenesses that will be imprinted
by such an optimized cross-matching scheme. It is possible that a
single cross-matching procedure is not suitable for all eROSITA science
projects, and that a number of individually tailored cross-matching
schemes will be required, depending on the patch on the sky.

The bulk of the X-ray sources in our study are stars and AGN, which are
intrinsically variable objects. However, we have made the simplifying
assumption throughout this work that variability (in luminosity and/or
in spectral energy distribution) of X-ray sources is not important for
the purposes of counterpart selection. This means that we do not take
account of extremely interesting, but difficult to handle scenarios
such as where an AGN that was bright at the epoch of its X-ray
detection (e.g. in ROSAT) has faded substantially (in all wavebands)
several years later when the measurement of its longer wavelength
counterpart (e.g. WISE or SDSS) was made
\citep[e.g. `changing-look'
QSOs;][]{LaMassa:2015cr,Merloni:2015oq, Runnoe:2016tg}. However, in
the future we will use AGN and stellar variability to our advantage
when selecting counterparts to eROSITA X-ray sources. With the
present (PTF/iPTF/ZTF\footnote{\url{http://www.ptf.caltech.edu/iptf}}\citealt{Rau:2009bs}; Catalina\footnote{\url{http://crts.caltech.edu/}}, Pan-STARRS\footnote{\url{https://panstarrs.stsci.edu/}}:
\citealt{Chambers:2016fv}; etc.) and forthcoming generation of optical time domain surveys, \citep[e.g. as
performed by the Large Synoptic Survey Telescope;][]{Gressler:2014fv}, every
potential optical counterpart to an X-ray source will also come with
robust measurements of optical variability. Such variability metrics,
which naturally separate AGN and stars from field galaxies, and can be
simply applied as an additional prior in \nway \citep[see, for example,][]{Budavari:2017ff}.

%--------------------------------------
%----------------------------------------------------

\section*{Acknowledgement}
The authors are grateful to the referee for the valuable input on the earlier version of this paper.
The authors are grateful to  M. J. Freyberg, T. Boller and F. Haberl for their help in understanding the 
old and new ROSAT catalogues.
MS and JB are grateful to G.Hasinger and T. Simm for testing and proving feedback on the various 
versions of \nway.
MS thanks F. Guglielmetti, A. Georgakakis and D. Coffey for various discussions over the years that 
helped in shaping the final work.
JB acknowledges support from FONDECYT Postdoctorados grant 3160439 and
the Ministry of Economy, Development, and Tourism's Millennium Science
Initiative through grant IC120009, awarded to The Millennium Institute
of Astrophysics, MAS.
MB acknowledges support from the FP7 Career Integration Grant ``eEASy?' (CIG 321913).
This publication makes use of data products from the Wide-field Infrared Survey Explorer, which is a 
joint project of the University of California, Los Angeles, and the Jet Propulsion Laboratory/California 
Institute of Technology, funded by the National Aeronautics and Space Administration.
This research has made use of data obtained from the 3XMM XMM-Newton serendipitous source catalogue 
compiled by the 10 institutes of the XMM-Newton Survey Science Centre selected by ESA.
This publication makes use of data products from the Two Micron All Sky Survey, which is a joint project 
of the University of Massachusetts and the Infrared Processing and Analysis Center/California Institute 
of Technology, funded by the National Aeronautics and Space Administration and the National Science 
Foundation.
This research has made use of data obtained from XMMSL2, the Second XMM-Newton Slew Survey Catalogue, 
produced by members of the XMM SOC, the EPIC consortium, and using work carried out in the context of 
the EXTraS project ("Exploring the X-ray Transient and variable Sky", funded from the EU's Seventh 
Framework Programme under grant agreement no. 607452).
This work has made use of data from the European Space Agency (ESA)
mission {\it Gaia} (\url{https://www.cosmos.esa.int/gaia}), processed by
the {\it Gaia} Data Processing and Analysis Consortium (DPAC,
\url{https://www.cosmos.esa.int/web/gaia/dpac/consortium}). Funding
for the DPAC has been provided by national institutions, in particular
the institutions participating in the {\it Gaia} Multilateral Agreement.
This research has made  use of the Vizier catalogue access tool, CDS, Strasbourg, France. The original 
description of the VizieR service was published in A\&AS 143, 23.
This publication makes use of TOPCAT \citep{Taylor:2005bd} and STILTS \citep{Taylor:2006ud} available at 
\url{http://www.starlink.ac.uk/topcat/ } and \url{http://www.starlink.ac.uk/stilts/}, respectively.

%\newpage

\appendix
\section{A brief history of the matching problem}
\label{A:intro}

In astrophysics a source can be characterized by its accurate position on the sky, its  redshift and its 
Spectral Energy Distribution (SED).
If the positional accuracy is not known at a sub-arc second precision, the source cannot be the target 
of a spectroscopy study, and/or multi wavelength data cannot be correctly assembled.
While sources that are identified in the Optical and Near-Infrared regime usually have the required 
precision, this is not the case for sources selected at shorter and longer wavelengths. 
For example in the Far-infrared bands, Herschel reaches 6-7$\as$ Point Spread Function (PSF) at  70$\mu
$m, increasing up to $\sim$13$\as$ at longer wavelength. Similarly, in X-ray the positional measurement 
error depends on the counts and spatially varying PSF and therefore is not constant between sources. 
Typical positional uncertainties go from up to $\approx3\as$ ({\it Chandra}), to 7$\as$ ({\it XMM}) but reach up to about 
29$\as$ for 95\% of the  of ROSAT sources in 2RXS, with the values increasing toward the periphery of 
the field of view, up to more than 1$\am$ in the extreme cases.
This low positional accuracy, together with the fact that sources with different SEDs and different redshift emit 
the bulk of their energy in different photometric bands, make it difficult to identifying with certainty 
the {\it same} source in different surveys. Additionally, the entire pairing process is done by means of 
catalogues which can differ in depth, technique  for  "source detection" (and definition thereof).
In the past, the data were so shallow that a simple cross match in coordinates between catalogues was 
enough for pairing correctly the sources. Now, we reach sources that are so faint that we must adopt a 
probabilistic approach.

 The most used technique is based on  the Likelihood Ratio (LR) method \citep{Sutherland:1992fv}.
Taking into account source number densities, coordinates (with relative errors) and magnitude 
distribution of the sources, the method estimates the ratio between the  likelihood that a given source 
from catalogue B is the correct counterpart to a source detected in a catalogue A, and the likelihood of  
being a source in the background. Different factors are then considered when computing the threshold 
above which the likelihood ratio assures a reliable association. The procedure is  repeated anew for the 
pairing between the catalogues A-C, A-D, etc. 
If catalogues are i) from images at similar wavelength and  ii) of sufficient depth, for most of the 
sources in A, the counterpart in catalogues B, C, D etc will be the same, while for a fraction of the 
sources further considerations based on the shape of the SED  need to be taken into account for the 
counterpart association.

Moving from a generic description to a specific application, let us focus from now on to the case of 
finding the correct counterpart to X-ray sources. The LR method has been successfully applied on surveys 
like  XMM-COSMOS  \citep{Brusa:2007fp, Brusa:2010lr}, CDFS \citep{Luo:2008kh, Xue:2011lr, Luo:2017fk}, 
{\it Chandra}-COSMOS \citep{Civano:2012ys, Marchesi:2016jw},  XXL \citep{Georgakakis:2017ys}, STRIPE-82X 
\citep[][]{LaMassa:2016cq, Tasnim-Ananna:2017gf},  AEGIS-X \citep{Georgakakis:2011dq, Nandra:2015uq} 
just to mention a few.
For  each of these surveys, the authors performed the  steps described above,  pairing X-ray to 
optical, to near-infrared  and to mid-infrared data, independently. Then, they  ranked the ancillary 
data available in order of reliability  (i.e. deep and higher resolution data first) for selecting the 
correct counterpart in those cases where the LR method does not  provided a unique solutions.
 
The Bayesian approach is increasingly favored by the entire community.
Contrary to the LR method that is data-driven, the Bayesian approach  uses a model for reference  
(prior) and thus can be applied also to small samples and areas. This is a strength of the method but  
but a frequent criticism is that 
the assumption of a {\it model distribution} might not represent the reality.
These criticisms are legitimate  in general but in the specific case of  finding 
the counterpart to X-ray detected sources they are somewhat outdated. In fact, deep {\it Chandra} and 
{\it XMM} surveys are  so advanced/extended that  reliable models of  magnitude distribution of the 
counterparts to sources detected up to a desired depth, can now  be constructed empirically. Another virtue of the Bayesian approach is that many priors can be adopted, each independent of the next. So we can adopt a Bayesian form for the probability of a sources to be the right 
counterpart based on its position, its magnitude, colour etc.

At the basis of many  Bayesian cross-matching algorithms  is the formalism  introduced by 
\citet{Budavari:2008qy}\footnote{However, the work does not correctly account for the sources that for 
physical reasons (e.g. due to the shape of the SED, redshift value)  are missed in some of the 
catalogues. This has been pointed out by many authors  \citep[e.g.][]{Roseboom:2009kx, Pineau:2011ff} }.
This enables simultaneously cross-matching of multiple catalogues and provides the
Bayes factor fror the astrometric measures.
This Bayes factor from the astrometry  is then combined with one (or more) related 
to physical properties. E.g., \citet{Roseboom:2009kx} search the right
counterparts to sub-millimeter sources by computing the photometric redshift and SED fitting of each 
source within a certain radius circle.

Independently from the adopted method, an additional difficulty arises when the goal is to find the 
counterparts to X-ray  surveys that cover over hundreds of square degrees \citep[e.g., eROSITA:][]{Merloni:2012uq}.
In this case, the multi-wavelength catalogues from where to draw  the correct identification 
will not be homogeneously covering the field, but rather a patchwork of different surveys/depths, thus 
effecting the actual magnitude distribution of the field sources and thus the determination of the real 
counterpart.\\

In view of these new challenges, we designed \nway, an algorithm based on two-steps Bayesian approach. 
In the following   we provide the complete description of the code and its application to test cases in 
COSMOS. The code is released, together with a detailed manual and a set of test data for training purposes. 
purposes.

\section{Matching Methodology}
\label{A:nway}

This section lays out in detail the computation \nway performs. Further details and clear explanations 
on the use of the \nway are presented in the manual and tutorial of the code, distributed via Github at 
\protect\url{https://github.com/JohannesBuchner/nway}.

The features of \nway include:
\begin{enumerate}
\item Matching of N catalogues simultaneously.
\item Computation of all combinatorially possible matches.
\item Consideration of partial matches across catalogues, i.e. the absence
 of counterparts in some catalogues.
\item Taking into account the positional uncertainties and the source number densities, computation of 
the probability of each possible match.
\item Computation of the probability that there is no match.
\item Incorporating magnitude, colour or other information about the sources
of interest, refining the match probabilities.
\end{enumerate}

This is done in several steps:
\begin{enumerate}
\item Finding combinatorially all possible matches. See Section~\ref{sec:cartesian-product}.
\item Computing each match probability from number densities, separation distances and positional errors 
alone, taking into account the chance of a random alignment. See Section~\ref{sec:match-distance}.
\item For each source of the primary catalogue (in the application from
this paper: for each the X-ray source), compute (a) the probability
that this source does not have a counterpart and (b), assuming this
source has a counterpart, compute the relative probability for each
possible match. See Section~\ref{sec:grouping}.
\item Refining the probabilities by additional prior information. See Section~\ref{sec:match-prior}.
\end{enumerate}
In \nway, only the first catalogue (\emph{primary catalogue}) has
a special role. For every entry in this catalogue, matches are sought
in the other catalogues. The entries in the primary catalogue must
come with an ID. All catalogues must contain RA, DEC, positional error
information, the size of the area of sky covered by the catalogue.
The latter information is used to compute the probability of a chance
alignment.

\subsection{Computing all possible matches}
\label{sec:cartesian-product}

\begin{figure}
\emph{Input:}

\begin{tabular}{|>{\centering}m{2cm}|c|>{\centering}p{2cm}|c|>{\centering}p{2cm}|}
\cline{1-1} \cline{3-3} \cline{5-5} 
Primary Catalogue &  & 2nd Catalogue &  & 3rd Catalogue\tabularnewline
\cline{1-1} \cline{3-3} \cline{5-5} 
x1 &  & b1 &  & c1\tabularnewline
x2 &  & b2 &  & c2\tabularnewline
... &  & ... &  & ...\tabularnewline
 &  &  &  & \tabularnewline
\cline{1-1} \cline{3-3} \cline{5-5} 
\end{tabular}

\emph{\vspace{0.1cm}
}

\emph{Output:}

\emph{\vspace{0.01cm}
}
\begin{tabular}{|>{\centering}m{1.5cm}>{\centering}p{1.5cm}>{\centering}p{1.5cm}>{\centering}p{1cm}||c}
\cline{1-4} 
\emph{\vspace{-0.1cm}
}

Primary Cat. Entry & 2nd Cat. Entry & 3rd Cat. Entry & \multicolumn{1}{>{\centering}p{1cm}|}{Proba-\\
bility} & \tabularnewline
\cline{1-4} 
x1 & b1& c1 & ... & \multirow{9}{*}{source x1}\tabularnewline
x1 & b1 & c2& ... & \tabularnewline
x1 & b1 & (none) & ... & \tabularnewline
x1 & b2 & c1 & ... & \tabularnewline
x1 & b2 & c2 & ... & \tabularnewline
x1 & b2 & (none) & ... & \tabularnewline
x1 & (none) & c1 & ... & \tabularnewline
x1 & (none) & c2 & ... & \tabularnewline
x1 & (none) & (none) & ... & \tabularnewline
x2 & ... & ... & \multicolumn{1}{>{\centering}p{1cm}|}{...} & source x2\tabularnewline
\end{tabular}

\caption{\label{fig:cartesian-product}All possible combinations of matches
from the input catalogues are combined into the output catalogue.
Each such match has a computed probability, either based on positions and number densities or 
additionally refined after the adoption of one or more priors. The matches are grouped by the primary 
catalogue entries (here: x1, x2). }
\end{figure}

%--------------------------------------

First, possible associations are found. Figure \ref{fig:cartesian-product}
shows that all possible associations between the input catalogues
are considered when building the output catalogue. For this, a hashing
procedure puts each object into HEALPix bins \citep{Gorski:2005dz}. The bin width $w$ is
chosen so that an association of distance $w$ is improbable, i.e.
much larger than the largest positional error. An object with coordinates
$\phi,\,\theta$ is placed in the bin corresponding to its coordinate, but also into
its neighbouring bins to avoid boundary effects. This is done for each catalogue separately.
Then, in each bin, the Cartesian product across catalogues (every
possible combination of sources) is computed. All associations are
collected across the bins and filtered to be unique. 
The hashing procedure adds very low effort $O(\sum_{i=1}^{k}N_{i})$
while the Cartesian product is reduced drastically to $O(N_{\text{bins}}\cdot\prod_{i=1}^{k}\frac{N_{i}}
{N_{bins}})$,
from a naive approach complexity of $O(\prod_{i=1}^{k}N_{i})$
All primary objects that have no associations past this step have
$P(\text{"any real association"}|D)=0$.

A problem arises when the secondary catalogues  have depths or resolution such that some of the  sources 
appear only in some of the catalogues . 
So we need to consider also pairing that do not include a source from the primary catalog. The 
computation becomes infeasible
very quickly as the number of catalogues reaches four or more, as demonstrated in \citet{Pineau:2017qy}.
 
\nway first considers as an initial list all possibilities
which have the primary catalogue source in an association. As shown
above, this includes associations where some catalogues do not participate.
The remaining sources are considered independent. Secondly, associations
across the unused catalogues are considered for each case. To do this
with low computational complexity, the additional associations considered
are those in the initial list, but with the primary catalogue source
removed. For instance, for the case of primary source x1 with the other
sources independent, x1-(none)-(none), the additional associations
to consider are x1-b1-c1, x1-b1-c2, x1-b2-c1,
x1-b2-c2, i.e. with the primary source removed, b1-c1,
b1-c2, b2-c1, b2-c2. The
highest distance-based posterior of these additional associations
is factored into the distance-based posterior of the association with
the primary source. In practice, this solves the problem of tight
unrelated associations (e.g. b2-c1), which, if not
considered otherwise, would unduly favor an association which includes
them (e.g. x1-b1-c1). If five or more catalogues are
matched, not only one but two additional simultaneous association
might need to be considered. The impact of our approximation then
depends on the application. Our choice of using the highest posterior
over all unrelated associations is expected to handle such many-catalogue
applications well. If however several groups of similar nature (e.g.,
an X-ray catalogue, two radio catalogues and three optical catalogues)
are to be matched, proceeding hierarchically may give better results
(e.g. first match the optical catalogues together). However, more
testing is needed in this area.

\subsection{Distance-based matching}
\label{sec:match-distance}

The second step is the computation of association probabilities using
the angular distances between counterparts. In the last step (Section
\ref{sec:grouping}), for each source in the primary catalogue these
probabilities from the various possible matches are combined. In the
end this gives the probability that this source does not have a counterpart
and, assuming this source has a counterpart, compute the relative
probability for each possible match. At this step however we first
compute the probability for a particular association (e.g. x1-b1-c1,
or x1-(none)-c2) to be actually the same object versus a
chance alignment of unrelated objects.

The probability of a given association is computed by comparing the
probability of a random chance alignment of unrelated objects (prior)
to the likelihood that the sources from the various catalogues are
in fact the same object. The prior is evaluated from the density of
each catalogue and their effective coverage. Varying depths between
the catalogues and different coverage can further reduce the fraction
of expected matches, which can be adjusted for with a user-supplied
incompleteness factor. The posterior for each association based on
the distances only is calculated (output column {\code dist\_post}). The
mathematical details of this computation be found in Section~\ref{sub:math-probability}.
This probability can be modified by additional information (see Section~\ref{sec:match-prior}).

\subsection{Grouping, Flagging and Filtering}
\label{sec:grouping}

In the final step, associations are grouped by the source from the
primary catalogue (in our example, the X-ray catalogue). The posterior
probabilities that this source has any real association and the relative
probability for each match are computed (output columns {\code p\_any}
and {\code p\_i} respectively). Section~\ref{sub:math-probability}
details this computation. To remove low-probability associations from
the output catalogue, the user parameter \code{-{}-min-prob} can
be used to specify a threshold. The best match is indicated with \code{match\_flag\textbf{=}1}
for each primary catalogue entry. Secondary, almost as good solutions
are marked with \code{match\_flag\textbf{=2}}\footnote{While there can be only one source from the 
secondary catalogues  with \code{match\_flag\textbf{=}1} per each source of the primary catalog, there 
can be many  that are flagged \code{match\_flag\textbf{=2}}}.
By default associations are flagged with \code{match\_flag\textbf{=2}} as soon as 
{\code p\_i}$_{\code{match\_flag=1}}$/{\code p\_i}$_{\code{match\_flag=2}}>0.5$, but the use can 
change the threshold with
the parameter \code{-{}-acceptable-prob}. All other associations
are marked with \code{match\_flag\textbf{=}0}.

In the output catalogue the last three columns (\code{p\_any}, \code{p\_i}, \code{match\_flag})
allow the user to identify sources with one solution, possible secondary
solutions, and to build final catalogues.

\subsection{Matching with additional prior information}
\label{sec:match-prior}

For many classes of sources, the Spectral Energy Distribution (SED)
provides additional hints which associations are likely real. For
instance, the WISE colour distribution is different for X-ray sources
than for other objects (demonstration in Section \ref{sec:prior1}).
A powerful feature of \nway is to take advantage of this additional
information to improve the matching. In particular \nway allows:
\begin{enumerate}
\item Multiple priors to be used from any of the input catalogues.
\item Arbitrary quantities can be used. Providing priors is not limited
to magnitude distributions, one can use any other discriminating information
(e.g. colours, morphology, variability, etc.). 
\item It is possible to input pre-constructed  information, or compute the
distributions from the catalogues themselves based on secure distance-only
matches (see Section \ref{sub:Auto-calibration}).
\end{enumerate}
Section \ref{sec:mag-priors} has the mathematical details and a comparison
to the Likelihood Ratio method \citep{Sutherland:1992fv}

\subsection{Probability for an individual association}
\label{sub:math-probability}

Let us consider the problem of finding counterparts to a primary catalogue
($i=1$), in our example for the X-ray source position catalogue.
Let each $N_{i}$ denote the number of entries for the catalogues
used, and $\nu_{i}=N_{i}/\Omega_{i}$ denote their respective source surface
density on the sky. 

If a counterpart is required to exist in each of the $k$ catalogues,
there are $\prod_{i=1}^{k}N_{i}$ possible associations. If we assume
that a counterpart might be missing in each of the matching catalogues,
there are $N_{1}\cdot\prod_{i=2}^{k}(N_{i}+1)$ possible associations.
This minor modification, negligible for $N_{i}\gg1$, is ignored in
the following for simplicity, but handled in the code.

If each catalogue covers the same area with some respective, homogeneous
source density $\nu_{i}$, the probability of a chance alignment on
the sky of  $k$ physically unrelated objects can then be written \citep[eq. 25]{Budavari:2008qy}
as 
\begin{equation}
P(H)=N_{1}/\prod_{i=1}^{k}N_{i}=1/\prod_{i=2}^{k}N_{i}=1/\prod_{i=2}^{k}\nu_{i}\Omega_{i}.
\end{equation}
Thus $P(H)$ is the prior probability of an association. The posterior
should strongly exceed this prior probability, to avoid false positives. 

To account for non-uniform coverage, $P(H)$ is modified by a ``prior
completeness factor'' $c$, which gives the expected fraction of
sources with reliable counterpart (due to only partial coverage of
the matching catalogues $\Omega_{i>1}\neq\Omega_{1}$, depth of the
catalogues and/or systematic errors in the coordinates). Our prior
can thus be written as 
\begin{equation}
P(H)=c/\prod_{i=2}^{k}\nu_{i}\Omega_{1}.\label{eq:prior}
\end{equation}

Bayes theorem connects the prior probability $P(H)$ to the posterior
probability $P(H|D)$, by incorporating information gained from the
observation data $D$ via 
\begin{equation}
P(H|D)\propto P(H)\times P(D|H).\label{eq:bayes}
\end{equation}

We now extend the approach of \citet{Budavari:2008qy}, to allow matches
where some catalogues do not participate in a match. Comparing A12
and A14 in \citet{Budavari:2008qy}, assuming that positions lie on the
celestial sphere and adopting the expansions developed in their Appendix
B, we can write down likelihoods. For a counterpart across $k$ catalogues,
we obtain:

\begin{equation}
P(D|H)=2^{k-1}\frac{\prod\sigma_{i}^{-2}}{\sum\sigma_{i}^{-2}}\exp\left\{-\frac{\sum_{i<j}\phi_{ij}
\sigma_{j}^{-2}\sigma_{i}^{-2}}{2\sum\sigma_{i}^{-2}}\right\} \label{eq:nwaylikelihood}
\end{equation}
The likelihood for the hypothesis where some catalogues do not participate
in the association has the appropriate terms in the products and sums
removed. Therefore, the likelihood is unity for the hypothesis that
there is no counterpart in any of the catalogues.

In comparison to our method, the method of \citet{Budavari:2008qy} only
compares two hypotheses for an association: either all sources belong
to the same object ($H_{1}$), or they are coincidentally aligned
($H_{0}$). In this computation each hypothesis test is run in isolation,
and relative match probabilities for a given source are not considered.
For completeness, we also compute the posterior of this simpler model
comparison:

\begin{eqnarray}
\frac{P(H_{1}|D)}{P(H_{0}|D)} & \propto & \frac{P(H_{1})}{P(H_{0})}\times\frac{P(D|H_{1})}{P(D|H_{0})}\\
B & = & \frac{P(D|H_{1})}{P(D|H_{0})}\\
P(H_{1}|D) & = & \left[1+\frac{1-P(H_{1})}{B\cdot P(H_{1})}\right]^{-1}\label{eq:assocPost}
\end{eqnarray}

The output column \code{dist\_bayesfactor} stores $\log B$, while
the output column \code{dist\_post} is the result of equation \ref{eq:assocPost}.
The output column \code{p\_single} is the same as \code{dist\_post},
but modified if any additional information is specified (see Section
\ref{sec:mag-priors}). As mentioned several times in the literature,
the \citet{Budavari:2008qy} approach does not include sources absent
in some of the catalogues, while the formulae we develop below incorporate
absent sources. This is similar in spirit to \citet{Pineau:2017qy},
although the statistical approach is different. We now go further
and develop counterpart probabilities.

\label{sub:Grouping-by-primary}The first step in catalogue inference
is whether the source has any counterpart (\code{p\_any}).The
posterior probabilities $P(H|D)$ are computed using Bayes theorem
(eq. \ref{eq:bayes}) with the likelihood (eq. \ref{eq:nwaylikelihood})
and prior (eq. \ref{eq:prior}) appropriately adopted for the number
of catalogues the particular association draws from. For each entry
in the primary catalogue, the posteriors of all possible associations
are normalized to unity, and $P(H_{0}|D)$, the posterior probability
of the no-counterpart hypothesis, i.e., no catalogue participates,
computed. From this we compute:

\begin{equation}
\code{p\_any}=1-P(H_{0}|D)/\sum_{i}P(H_{i}|D)\label{eq:post-any}
\end{equation}
If \code{p\_any} is low, this indicates that there is little evidence
for any of the considered, combinatorially possible associations,
except for the no-association case. The output column \code{p\_any}
is the result of equation \ref{eq:post-any}.

If \code{p\_any}$\approx1$, there is strong evidence for at least
one of the associations to another catalogue. To compute the relative
posterior probabilities of the options, we re-normalize with the no-counterpart
hypothesis, $H_{0}$, excluded:

\begin{equation}
\code{p\_i}=P(H_{i}|D)/\sum_{i>0}P(H_{i}|D)\label{eq:post-assoc}
\end{equation}
If a particular association has a high $p_{i}$, there is strong evidence
that it is the true one, out of all present options. The output column
\code{p\_i} is the result of equation \ref{eq:post-assoc}.

A ``very secure'' counterpart could be defined by the requirement
\code{p\_any}$>$95\%and \code{p\_i}$>$95\%, for example. However, it is useful
to run simulations to understand the rate of false positives. Typically,
much lower thresholds are acceptable, with the threshold (dictated by the scientific applications) being 
a compromise between purity and completeness of the sample.

\subsection{Magnitudes, colours and other additional information}
\label{sec:mag-priors}

Specific classes of astronomical objects show distinct distribution on colour, magnitude  or other 
parameters, compared with the field population distributions. This can be exploited for finding the 
correct counterparts.
 Previous works \citep[e.g.][]{Ciliegi:2003fc, Ciliegi:2005rq,Brusa:2005kx,Brusa:2007fp}
have modified the likelihood ratio coming from the angular distance
$f(r)$ information (likelihood ratio method, \citealp{Sutherland:1992fv})
by a factor:

\begin{equation}
LR=\frac{q(m)}{n(m)}\times f(r)
\end{equation}
Here, $q(m)$ and $n(m)$ are associated with the magnitude distributions
of source (e.g. X-ray sources) and background objects respectively, but additionally contain sky density
contributions.

This idea can be put on solid footing within the Bayesian framework.
Here, two likelihoods are combined, by simply considering two independent
observations, namely one for the positions, $D_{\phi}$, and one for
the magnitudes $D_{m}$. The likelihood thus becomes

\begin{eqnarray}
P(D|H) & = & P(D_{\phi}|H)\times P(D_{m}|H)\\
 & = & P(D_{\phi}|H)\times\frac{\bar{q}(m)}{\bar{n}(m)},
\end{eqnarray}
with $\bar{q}(m)$ and $\bar{n}(m)$ being the probability that a
target (e.g. X-ray) source or a generic source in the field has magnitude $m$, respectively. 

\nway stores the modifying factor, $P(D_{m}|H)$,
in \code{bias\_{*}} output columns, one for each column giving a
magnitude, colour, or other distribution. This modifying factor is
however renormalized so that $P(D_{m}|H)=\frac{\bar{q}(m)}{\bar{n}(m)}/\int\frac{\bar{q}(m')}{\bar{n}
(m')}\bar{n}(m')dm'$,
which makes $P(D|H)=P(D_{\phi}|H)$ when $m$ is unknown. In that
case, $m$ is marginalized over its distribution in the general population,
i.e. $\int P(D_{m}|H)\,\bar{n}(m')\,dm$. This has the benefit that
when $m$ is unknown, the modifying factor is unity and the probabilities
remain unmodified.

For completeness, we mention the fully generalized case. This is attained
when an arbitrary number of photometry bands are considered, each
consisting of a magnitude measurement $m$ and measurement uncertainty
$\sigma_{m}$:

\begin{equation}
P(D_{m}|H)=\prod\frac{\int_{m}\bar{q}(m)\,p(m|D_{m})\,dm}{\int_{m}\bar{n}(m)\,p(m|D_{m})\,dm}
\end{equation}
Here, $p(m|D_{m})$ would refer to a Gaussian error distribution with
mean $m$ and standard deviation $\sigma_{m}$. This is convolved
with the distribution properties. Alternatively, $p(m|D_{m})$ can
also consider upper limits. However, such options are not yet implemented
in \nway. Instead, we recommend removing magnitude values with large
uncertainties (setting them to -99).

\subsubsection{Auto-calibration}
\label{sub:Auto-calibration}

The probability distributions $\bar{n}(m)$ and $\bar{q}(m)$ can
be taken from other observations by computing the normalized magnitude\footnote{We make the examples 
using magnitudes, but everything will work the same using any other parameter like  colours, morphology, 
variability etc.} histograms
of the overall population and the target sub-population (e.g. X-ray
sources). In \nway, the distributions $\bar{q}(m)$ and $\bar{n}(m)$
can be provided as an ASCII table, with the columns describing the
bin edges, the frequency of the target population (in our example,
X-ray sources) and the frequency of the field population (sources
that are not X-ray sources, at the depth of the catalogue).

Under certain approximations and assumptions, these histograms can
also be computed during the catalogue matching procedure 
 used for the weighting on the fly and saved for future further use.
For example, one could perform the distance-based
matching procedure laid out above, and compute a magnitude histogram
of the secure counterparts as an approximation for $\bar{q}(m)$ and
a histogram of ruled out counterparts for $\bar{n}(m)$. While the
weights $\bar{q}(m)/\bar{n}(m)$ may strongly influence the probabilities
of the associations for a single object, the bulk of the associations
will be dominated by distance-weighting. One may thus assume that
the $\bar{q}(m)$ and $\bar{n}(m)$ are computed with and without
applying the magnitude weighting are the same, which is true in practice.
When differences are noticed, they will only strengthen $\bar{q}(m)$,
and the procedure may be iterated.

In \nway auto mode, the histogram
$\bar{q}(m)$ is constructed using sources with \code{dist\_post>0.9}
(safe matches), and $\bar{n}(m)$ with \code{dist\_post<0.01} (safe
non-matches). 
When these "self constructed priors" are used, the breaks of the
histogram bins are computed adaptively based on the empirical cumulative
distribution found. Because the histogram bins are usually larger
than the magnitude measurement uncertainty, the latter is currently
not considered. The adaptive binning creates bin edges based on the
number of objects, and is thus independent of the chosen scale (magnitudes,
flux). Thus the method is not limited to magnitudes, but can be used
for virtually any other known object property (colours, morphology,
variability, etc.), as demonstrated in the main body of this paper.

\section{Testing NWAY on COSMOS}
\label{A:TestCosmos}

The COSMOS field  \citep{Scoville:2007rw} offers the ideal test bench, covering a relatively large  
area with homogeneous and deep observations in many bands. In particular, the field has been 
observed with {\it XMM-Newton} \citep{Hasinger:2007dn, Cappelluti:2007fj}, and its reliable association  
to the I-band CFHT/Megacam catalogue \citet{McCracken:2007ee} via LR is presented in \citet{Brusa:2007fp}.
Successively, \citet{Brusa:2010lr} improved on the first associations using also the near-infrared \citep{McCracken:2010pj} and  the mid-infrared \citep{Sanders:2007qd, Ilbert:2010qy} catalogues. Each catalogue was used independently and the counterparts were chosen via LR and visually 
inspected. 

More recently, for the same area, deeper and homogeneous observations from  {\it Chandra} became 
available \citep{Elvis:2009kx, Civano:2012ys, Civano:2016kb, Marchesi:2016jw} so  that the  XMM-COSMOS 
associations have been successively validated/changed on the basis of the smaller positional 
uncertainties of the {\it Chandra} X-ray data\footnote{The user should refer to \citet{Marchesi:2016jw} 
for details about the comparison between XMM-/{\it Chandra}- COSMOS detections.} ($\sim$0.5$\as$ vs. $
\sim$2$\as$ for XMM, averaging over the entire FOV).

In the following two sections we describe the successful application of \nway to the XMM-COSMOS field,  
first using only one optical catalogue. We then repeat the association using simultaneously the  
optical and IRAC catalogues. We show how the associations and the key \nway parameters {\code p\_i} and 
{\code p\_any} change in the two applications.
The optical and IRAC catalogues are the original ones used by \citet{Brusa:2010lr}. They are released 
with \nway  and described in the manual so that a curious reader can practice with the code.

 \subsection{\nway Success rate}
 
The XMM-COSMOS catalogue of multi-wavelength counterparts presented in  \citet{Brusa:2010lr} included 
1822 sources, 1797 of which  are isolated\footnote{i.e. 25 sources correspond to two or more {\i 
Chandra} detections}. We focus  here on the 1281(128) isolated {\it XMM}-COSMOS sources with the 
original confirmed(changed) association after using {\it Chandra} data. 
  
We extracted from the  catalogue the identification number,  the X-ray 
coordinates and corresponding positional errors of the 1409 (1281+128) isolated  sources. The mean 
positional error of the sample is 1.8$\as$ with a minimum value of 0.1$\as$ to a max of 7.33$\as$. 
Similarly, we extracted from  the optical \citep {McCracken:2007ee}  and IRAC \citep{Sanders:2007qd, Ilbert:2010qy} catalogues
the identification numbers, the coordinates and the magnitude in the optical and 
3.6 $\mu m$ bands. We assumed, as in \citet{Brusa:2007fp}, a constant positional error of 0.1$\as$ and 
0.5$\as$ for the two catalogues, respectively.
  
First, we ran \nway with the {\it XMM} and optical catalogues in mode "auto" (see \S\ref{sub:Auto-calibration}).
Although we know that for this sample the actual counterparts are within 8$\as$ from the 
X-ray positions, we searched for a counterpart within a radius of 20$\as$ in order to avoid any bias in 
the result. 
In the 96\% (1231/1281) of the cases  \nway assigned the same  counterpart\footnote{For this test we 
consider as counterpart the source with the highest {\code p\_any}{\code p\_i} within each circle} as in 
\citet{Brusa:2007fp}. In addition, of the 128 sources for which the counterpart  has changed thanks to 
the higher resolution of {\it Chandra}, \nway recovered correctly (and independently) 55 of them (i.e. 
43\%). 

In the second test, we run \nway again in mode "auto", but this time pairing simultaneously the {\it 
XMM} catalogue to the  optical and IRAC catalogues. Intuitively, increasing the number of priors the 
number of correct associations should increase. At the same time, the number of matches due to chance 
association should decrease. In particular, a second prior will reinforce the probability that a source 
is the correct counterpart, or, provide an alternative, better counterpart.
 In fact in this second application we recovered correctly 1250/1281 (97.6\%) sources. Of the 128 
sources that change counterpart after {\it Chandra} observations, we  recovered correctly 65 (50.8\%) of 
them, without any additional information. The new sources were either very faint  or completely missed 
in optical catalog. 
 
\subsection{\nway parameters behaviors}

As discussed when describing the code, \nway provides the quantities {\code p\_any} and \code{p\_i}
that can help in assessing the reliability of  an association. 
The first parameter indicates what is the probability that an X-ray sources has at least a reliable 
counterpart among the possible associations, behaving as the prior. Low {\code p\_any} indicates that 
either the prior is not able to disentangle between possible counterparts, or that none of the possible 
counterparts behave as the prior.
The second parameter, \code{p\_i} indicates what is the probability for a given source to be the correct 
counterpart among the possible associations to an X-ray source.

\begin{figure*}
\label{Fig:xmm_opt_irac}

\includegraphics[width=8.0cm]{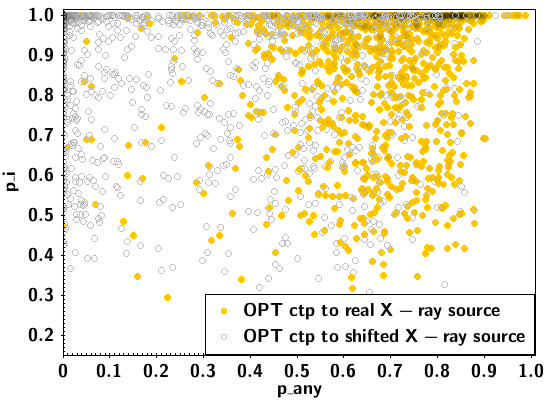}\includegraphics[width=8.0cm]{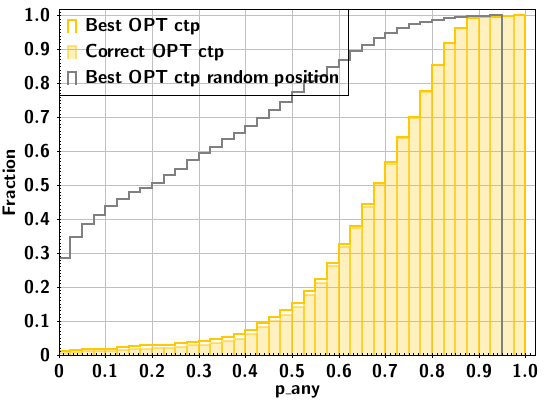}
\includegraphics[width=8.0cm]{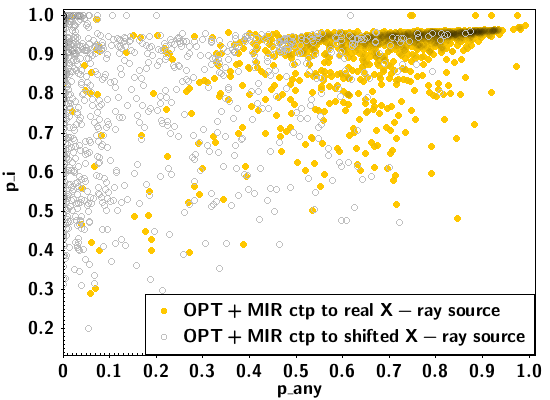}\includegraphics[width=8.0cm]{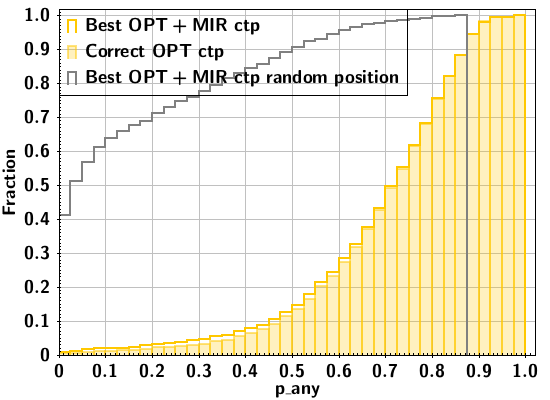}

\caption{{\bf Top left:} \code{p\_i}  vs. \code{p\_any} distribution for the correct counterparts to the 
{\it XMM}-COSMOS sources (yellow) and for the candidate counterpart to the same sources after 
randomizing their position (grey), using only the optical catalog.{\bf Top right:} Cumulative 
distribution of \code{p\_any} for the actual sample of correct counterparts (solid yellow)  and for the 
counterparts to the randomized X-ray sources (solid grey). The dashed yellow line represent the 
distribution, including also the 26 sources for which \nway fails in identifying the correct 
counterparts. These sources have all  \code{p\_any} below 0.5 (after that value the two yellow curves 
coincide).{\bf Bottom left:} and {\bf Bottom right:} As above, but this time using simultaneously a 
prior in Optical and one in mid-infrared.}
\end{figure*}

In Figure~\ref{Fig:xmm_opt_irac} we show {\code p\_i} versus {\code p\_any} for the optical prior (OPT, 
top-left panel) and for the optical+IRAC prior (OPT+MIR, bottom-left panel.). In Addition we show the 
respective cumulative distribution of {\code p\_any} (right panels).
Additionally, we plot in grey the same parameters as before computed for random associations. This was 
obtained by applying \nway to the same catalogues, but after randomizing the position of the X-ray 
sources by shifting by 1$\am$ their Declination. 

From the top plots we can see that the distribution of {\code p\_any} for the random position, 
concentrate at  low values while the {\code p\_any} for the real sources peak at high values. For example 
{\code p\_any}$_{real}>0.6$ for 80\% of the counterpart to real X-ray sources while only 0.09\% of the 
counterparts to randomized X-ray sources have such high {\code p\_any}\footnote{Note that we cannot 
exclude that these 0.09\% sources are the counterpart to real X-ray sources that are fainter that the 
depth of our survey}.

The term  \code{p\_i} is the combination of two terms, one related to the pure positional match and density of the sources and one 
related to the prior. 
If the number density of sources in the optical catalogue is high, there can always be a possible 
counterpart due to chance association, for the randomized X-ray sources. For this reason, more than 40\% 
of the possible counterparts to the randomized X-ray source have  \code{p\_i}$>$0.8. 
Only coupling {\code p\_i} with {\code p\_any} we can find out the actual nature of the counterpart.

The situation changes noticeably in the bottom panels of Figure~\ref{Fig:xmm_opt_irac}, where not one 
but two priors (one in Optical and one in mid-infrared) are simultaneously considered. Here, the 
distribution of \code{p\_i} and \code{p\_any} is similar to the previous case, while for the counterpart 
to actual X-ray sources, both parameters peak at high values. Again, it is noteworthy that 99.9\% of the counterparts were correctly
identified already with only one prior. The additional prior 
just increased  \code{p\_i} indicating how the real counterparts clearly stand up from the field 
distribution. Intuitively, adding a third prior would reduce even strongly the possibility that a 
counterpart is selected due to chance association.

Finally, an important point to stress is that while the original work on the {\it XMM} and {\it Chandra} 
associations took months and an additional visual inspection was necessary, the reliable results 
presented here for \nway were obtained in less than 5 minutes with a single 2700MHz CPU without any 
filtering or inspection.

%--------------------

\bibliographystyle{mnras} 
\bibliography{newlibrary.bib} 

%\bibliography{}
%\begin{thebibliography}{}
%\nocite{*}
%\printbibliograph
%\end{thebibliography}

 %%%%%%%%%%%%%%%%%%%%%%%%%%%%%%%%%%%%%%%%%%%%%%%%%%

% Don't change these lines
\bsp	% typesetting comment
\label{lastpage}
\end{document}